\documentclass[runningheads,a4paper]{llncs}

\usepackage{logic9-thm}

\usepackage{listings}
\usepackage{amsmath,wasysym}
\usepackage{multicol}
\usepackage{algorithm, algorithmic}

\usepackage{tikz}
\usetikzlibrary{automata,fit,positioning}

\newcommand{\xra}{\xrightarrow}

\newcommand{\Closure}{\operatorname{\mathit{Closure}}}

\newcommand{\tuple}[1]{\ensuremath{(#1)}}
\newcommand{\Rpos}[0]{\ensuremath{\mathbb{R}_{\geq 0}}}

\renewcommand{\Nat}[0]{\ensuremath{\mathbb{N}}}
\newcommand{\CC}[0]{\ensuremath{\Phi}} % clock-contraints
\newcommand{\val}[0]{\ensuremath{\nu}} % valuation
\newcommand{\vali}[0]{\ensuremath{\mathbf{0}}} % valuation: each x to 0

\newcommand{\reset}[1]{\ensuremath{[#1]}}

\newcommand{\Alu}{\mathfrak{a}_{\preccurlyeq LU}}
\newcommand{\Alui}{\mathfrak{a}_{\preccurlyeq_{L_1U_1}}}
\newcommand{\Alup}{\mathfrak{a}_{\preccurlyeq_{L_1'U_1'}}}
\newcommand{\lu}{\preccurlyeq_{LU}}

\newcommand{\Extra}{\mathit{Extra}}
\newcommand{\elup}[1]{\Extra^+_{LU}(#1)}
\newcommand{\Repeat}[2]{\noindent{$\blacktriangleright$\textsc{\textbf{#1~\ref{#2}.\
      }}}}

\newcommand{\floor}[1]{\lfloor #1 \rfloor}
\newcommand{\ceil}[1]{\lceil #1 \rceil}

\lstdefinelanguage{algo}{%
  morekeywords={function,algorithm,push,pop,top,for,all,and,or,if,then,else,repeat,until,while,do,report,return,such,that,each,add,call,exit,let}
}

\begin{document}

 \title{Using non-convex approximations for efficient analysis of timed
   automata}

\author{F. Herbreteau\inst{1}, D. Kini\inst{2},
  B. Srivathsan\inst{1} and I. Walukiewicz\inst{1}}
\institute{
Universit\'{e} de Bordeaux, IPB, Universit\'{e} Bordeaux 1, CNRS,
LaBRI UMR5800\\
LaBRI B\^{a}t A30, 351 crs Lib\'{e}ration, 33405 Talence, France
\and
Indian Institute of Technology Bombay, Dept. of Computer Science and Engg.,
\\ Powai, Mumbai 400076, India\\
}
\maketitle

\begin{abstract}
  The reachability problem for timed automata asks if there exists a
  path from an initial state to a target state. The standard
  solution to this problem involves computing the zone graph of the
  automaton, which in principle could be infinite. In order to make
  the graph finite, zones are approximated using an extrapolation
  operator. For reasons of efficiency in current algorithms
  extrapolation of a zone is always a zone; and in particular it
  is convex.

  In this paper, we propose to solve the reachability problem without
  such extrapolation operators. To ensure termination, we provide an
  efficient algorithm to check if a zone is included in the so
  called region closure of another. Although theoretically better,
  closure cannot be used in the standard algorithm since a closure of
  a zone may not be convex.

  An additional benefit of the proposed approach is that it permits to
  calculate approximating parameters on-the-fly during exploration of
  the zone graph, as opposed to the current methods which do it by a
  static analysis of the automaton prior to the exploration. This
  allows for further improvements in the algorithm. Promising
  experimental results are presented.
\end{abstract}

% Files
\section{Introduction}
\label{sec:introduction}

Timed automata~\cite{Alur:TCS:1994} are obtained from finite automata
by adding clocks that can be reset and whose values can be compared
with constants. The crucial property of timed automata is that their
reachability problem is decidable: one can check if a given target
state is reachable from the initial state.  Reachability algorithms
are at the core of verification tools like
Uppaal~\cite{Behrmann:QEST:2006} or RED~\cite{Wang:STTT:2004}, and are
used in industrial case
studies~\cite{HSLL:RTSS:1997,BBP:IJPR:2004}. The standard solution
constructs a search tree whose nodes are approximations of zones.  In
this paper we give an efficient algorithm for checking if a zone is
included in an approximation of another zone. This enables a
reachability algorithm to work with search trees whose nodes are just
unapproximated zones. This has numerous advantages: one can use
non-convex approximations, and one can compute approximating
parameters on the fly.

The first solution to the reachability problem has used
\emph{regions}, which are equivalence classes of clock
valuations. Subsequent research has shown that the region abstraction
is very inefficient and an other method using \emph{zones} instead of
regions has been proposed. This can be implemented efficiently using
DBMs~\cite{Dill:AVMFSS:1989} and is used at present in almost all
timed-verification tools. The number of reachable zones can be
infinite, so one needs an abstraction operator to get a finite
approximation. The simplest is to approximate a zone with the set of
regions it intersects, the so called \emph{closure} of a
zone. Unfortunately, the closure may not always be convex and no
efficient representation of closures is known. For this reason
implementations use another convex approximation that is also based on
(refined) regions.

We propose a new algorithm for the reachability problem using closures
of zones. To this effect we provide an efficient algorithm for
checking whether a zone is included in a closure of another zone. In
consequence we can work with non-convex approximations without a need
to store them explicitly.

Thresholds for approximations are very important for efficient
implementation. Good thresholds give substantial gains in time and
space. The simplest approach is to take as a threshold the maximal
constant appearing in a transition of the automaton. A considerable
gain in efficiency can be obtained by analyzing the graph of the
automaton and calculating thresholds specific for each clock and state
of the automaton~\cite{Behrmann:TACAS:2003}. An even more efficient
approach is the so called LU-approximation that distinguishes between
upper and lower bounds~\cite{Behrmann:STTT:2006}. This is the method
used in the current implementation of UPPAAL. We show that we can
accommodate closure on top of the LU-approximation at no extra cost.

Since our algorithm never stores approximations, we can compute
thresholds on-the-fly.  This means that our computation of thresholds
does not take into account unreachable states. In consequence in some
cases we get much better LU-thresholds than those obtained by static
analysis.  This happens in particular in a very common context of
analysis of parallel compositions of timed automata.

\subsection*{Related work}

The topic of this paper is approximation of zones and efficient
handling of them. We show that it is possible to use non-convex
approximations and that it can be done efficiently. In particular, we
improve on state of the art
approximations~\cite{Behrmann:STTT:2006}. Every forward algorithm
needs approximations, so our work can apply to tools like RED or
UPPAAL.

Recent work~\cite{Morbe:CAV:2011} reports on backward analysis
approach using general linear constraints. This approach does not use
approximations and relies on SMT solver to simplify the constraints.
Comparing forward and backward methods would require a substantial
test suite, and is not the subject of this paper.

\subsection*{Organization of the paper}

The next section presents the basic notions and recalls some of their
properties.  Section~\ref{sec:inclusion} describes the new algorithm
for efficient inclusion test between a zone and a closure of another
zone. The algorithm constructing the search tree and calculating
approximations on-the-fly is presented in Section~\ref{sec:lazy}.
Some results obtained with a prototype implementation are presented in
the last section. All missing proofs are presented in the full version
of the paper~\cite{Herbreteau:FSTTCS:2011:extended}.

%%% Local Variables: 
%%% mode: latex
%%% TeX-master: "m"
%%% End: 

% LocalWords:  reachability Kronos
 
%%
\section{Preliminaries}

\subsection{Timed automata and the reachability problem}

Let $X$ be a set of clocks, i.e., variables that range over $\Rpos$,
the set of non-negative real numbers. A \emph{clock constraint} is a
conjunction of constraints $x\# c$ for $x\in X$,
$\#\in\{<,\leq,=,\geq,>\}$ and $c\in \Nat$, e.g. $(x \le 3 \wedge y >
0)$. Let $\CC(X)$ denote the set of clock constraints over clock
variables $X$.  A \emph{clock valuation} over $X$ is a function
$\val\,:\,X\rightarrow\Rpos$. We denote $\Rpos^X$ the set of clock
valuations over $X$, and $\vali$ the valuation that associates $0$ to
every clock in $X$. We write $\val\sat \phi$ when $\val$ satisfies
$\phi\in \CC(X)$, i.e. when every constraint in $\phi$ holds after
replacing every $x$ by $\val(x)$. For $\d\in\Rpos$, let $\val+\d$ be
the valuation that associates $\val(x)+\d$ to every clock $x$. For
$R\subseteq X$, let $\reset{R}\val$ be the valuation that sets $x$ to
$0$ if $x\in R$, and that sets $x$ to $\val(x)$ otherwise.

A \emph{Timed Automaton (TA)} is a tuple $\Aa=\tuple{Q,q_0,X,T,\Acc}$
where $Q$ is a finite set of states, $q_0\in Q$ is the initial state,
$X$ is a finite set of clocks, $\Acc\subseteq Q$ is a set of accepting
states, and $T\,\subseteq\, Q\times\CC(X)\times 2^X \times Q$ is a
finite set of transitions $\tuple{q,g,R,q'}$ where $g$ is a
\emph{guard}, and $R$ is the set of clocks that are \emph{reset} on
the transition. An example of a TA is depicted in
Figure~\ref{fig:ta-A4}. The class of TA we consider is commonly known
as diagonal-free TA since clock comparisons like $x-y\leq 1$ are
disallowed. Notice that since we are interested in state reachability,
considering timed automata without state invariants does not entail
any loss of generality. Indeed, state invariants can be added to
guards, then removed, while preserving state reachability.

A \emph{configuration} of $\Aa$ is a pair $(q,\val)\in
Q\times\Rpos^X$; $(q_0,\vali)$ is the \emph{initial configuration}. We
write $\tuple{q,\val}\xrightarrow{\d,t} \tuple{q',\val'}$ if there
exists $\d\in\Rpos$ and a transition $t=(q,g,R,q')$ in $\Aa$ such that
$\val+\d\sat g$, and $\val'=\reset{R}\val$. Then $\tuple{q',\val'}$ is
called a \emph{successor} of $(q,\val)$. A \emph{run} of $\mathcal{A}$
is a finite sequence of transitions:
$\tuple{q_0,\val_0}\xrightarrow{\d_0,t_0}
\tuple{q_1,\val_1}\xrightarrow{\d_1,t_1} \cdots \tuple{q_n,\val_n}$
starting from $\tuple{q_0,\val_0}=\tuple{q_0,\vali}$.

A run is \emph{accepting} if it ends in a configuration
$\tuple{q_n,\val_n}$ with $q_n\in \Acc$. The \emph{reachability
  problem} is to decide whether a given automaton has an accepting
run. This problem is known to
be~\PSPACE-complete~\cite{Alur:TCS:1994,Courcoubetis:FMSD:1992}.

\subsection{Symbolic semantics for timed automata}
\label{sec:symbolic}

The reachability problem is solved using so-called symbolic semantics.
It considers sets of (uncountably many) valuations instead of
valuations separately. A \emph{zone} is a set of valuations defined
by a conjunction of two kinds of constraints: comparison of difference
between two clocks with an integer like $x-y\# c$, or comparison of a
single clock with an integer like $x\# c$, where
$\#\in\{<,\leq,=,\geq,>\}$ and $c\in\Nat$. For instance $(x-y\geq
1)\land(y<2)$ is a zone. The transition relation on valuations is
transferred to zones as follows. We have $(q,Z)\xra{t} (q',Z')$ if
$Z'$ is the set of valuations $\val'$ such that $(q,\val)\xra{\d,t}
(q',\val')$ for some $\val\in Z$ and $\d\in\Rpos$. The node $(q',Z')$
is called a successor of $(q,Z)$. It can be checked that if $Z$ is a
zone, then $Z'$ is also a zone.

The \emph{zone graph} of $\Aa$, denoted $ZG(\Aa)$, has nodes of the
form $(q,Z)$ with initial node $(q_0,\{\vali\})$, and edges defined as
above. Immediately from the definition of $ZG(\Aa)$ we infer that
$\Aa$ has an accepting run iff there is a node $(q,Z)$ reachable in
$ZG(\Aa)$ with $q\in Acc$.

Now, every node $(q,Z)$ has finitely many successors: at most one
successor of $(q,Z)$ per transition in $\Aa$. Still a reachability
algorithm may not terminate as the number of reachable nodes in
$ZG(\Aa)$ may not be finite~\cite{Daws:TACAS:1998}. The next step is
thus to define an abstract semantics of $\Aa$ as a finite graph. The
basic idea is to define a finite partition of the set of valuations
$\Rpos^X$. Then, instead of considering nodes $(q,S)$ with set of
valuations $S$ (e.g. zones $Z$), one considers a union of the parts of
$\Rpos^X$ that intersect $S$. This gives the finite abstraction.

Let us consider a \emph{bound function} associating to each clock $x$
of $\Aa$ a bound $\a_x\in\Nat$. A \emph{region}~\cite{Alur:TCS:1994}
with respect to $\a$ is the set of valuations specified as follows:
\begin{enumerate}
\item for each clock $x\in X$, one constraint from the set:

  $\{x = c ~\mid~ c=0, \dots, \a_x \} \cup \{c - 1 < x < c ~\mid~ c =
  1, \dots, \a_x \} \cup \{x > \a_x \}$
  
  \smallskip
\item for each pair of clocks $x,y$ having interval constraints: $c-1<
  x <c$ and $d-1< y< d$, it is specified if $fract(x)$ is less than,
  equal to or greater than $fract(y)$.
\end{enumerate}
It can be checked that the set of regions is a finite partition of
$\Rpos^X$.

The \emph{closure abstraction} of a set of valuations $S$, denoted
$\Closure_\a(S)$, is the union of the regions that intersect
$S$~\cite{Bouyer:FMSD:2004}. A simulation graph, denoted
$SG_\a(\Aa)$\label{SG:description}, has nodes of the form $(q,S)$
where $q$ is a state of $\Aa$ and $S\subseteq \Rpos^X$ is a set of
valuations. The initial node of $SG_\a(\Aa)$ is
$(q_0,\{\vali\})$. There is an edge $(q,S)\xra{t}
(q',\Closure_\a(S'))$ in $SG_\a(\Aa)$ iff $S'$ is the set of
valuations $\val'$ such that $(q,\val)\xra{\d,t} (q',\val')$ for some
$\val\in S$ and $\d\in\Rpos$. Notice that the reachable part of
$SG_\a(\Aa)$ is finite since the number of regions is finite.

The definition of the graph $SG_\a(\Aa)$ is parametrized by a bound
function $\a$. It is well-known that if we take $\a_{\Aa}$ associating
to each clock $x$ the maximal integer $c$ such that $x\# c$ appears in
some guard of $\Aa$ then $SG_\a(\Aa)$ preserves the reachability
properties.
\begin{theorem}{\cite{Bouyer:FMSD:2004}}\label{thm:symbolic-zone-graph}
  $\Aa$ has an accepting run iff there is a reachable node $(q,S)$ in
  $SG_\a(\Aa)$ with $q\in Acc$ and $\a_{\Aa}\leq \a$.
\end{theorem}

For efficiency it is important to have a good bound function $\a$. The
nodes of $SG_\a(\Aa)$ are unions of regions. Hence the size of
$SG_\a(\Aa)$ depends on the number of regions which is
$\Oo\big(|X|!.2^{|X|}.\prod_{x\in X}
(2.\a_x+2)\big)$~\cite{Alur:TCS:1994}. It follows that smaller values
for $\a$ yield a coarser, hence smaller, symbolic graph
$SG_\a(\Aa)$. Note that current implementations do not use closure but
some convex under-approximation of it that makes the graph even
bigger.

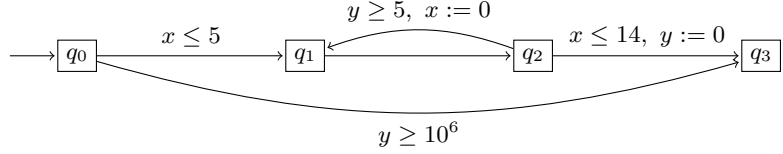
\begin{figure}[t]
  \centering
  \begin{tikzpicture}[shorten >=1pt,node distance=3cm,on
    grid,auto,every node/.style={shape=rectangle}]
    \node[draw] (q0) {\footnotesize $q_0$};
    \node[draw] (q1) [right=of q0] {\footnotesize $q_1$};
    \node[draw] (q2) [right=of q1] {\footnotesize $q_2$};
    \node[draw] (q3) [right=of q2] {\footnotesize $q_3$};
    \node (fakeinit) [node distance=1cm,left=of q0] {};
    \begin{scope}[->]
      \draw (fakeinit) edge (q0);
      \draw (q0) edge node {\footnotesize $x\leq 5$} (q1);
      \draw (q1) edge (q2);
      \draw (q2) edge [bend right=20] node [swap] {\footnotesize
        $y\geq 5,\ x:=0$} (q1);
      \draw (q2) edge node {\footnotesize $x\leq 14,\ y:=0$} (q3);
      \draw (q0) edge [bend right=16] node[swap] {\footnotesize $y\geq
        10^6$} (q3);
    \end{scope}
  \end{tikzpicture}
  \caption{Timed automaton $\Aa$.}
  \label{fig:ta-A4}
\end{figure}

It has been observed in~\cite{Behrmann:TACAS:2003} that instead of
considering a global bound function $\a_{\Aa}$ for all states in
$\Aa$, one can use different functions in each state of the automaton.
Consider for instance the automaton $\Aa$ in
Figure~\ref{fig:ta-A4}. Looking at the guards, we get that $\a_x=14$
and $\a_y=10^6$. Yet, a closer look at the automaton reveals that in
the state $q_2$ it is enough to take the bound $\a_y(q_2)=5$. This
observation from~\cite{Behrmann:TACAS:2003} points out that one can
often get very big gains by associating a bound function $\a(q)$ to
each state $q$ in $\Aa$ that is later used for the abstraction of
nodes of the form $(q,\Closure_{\a(q)}(S))$. In op.\ cit.\ an
algorithm for inferring bounds based on static analysis of the
structure of the automaton is proposed. In Section~\ref{sec:bounds} we
will show how to calculate these bounds on-the-fly during the
exploration of the automaton's state space.

%%% Local Variables: 
%%% mode: latex
%%% TeX-master: "m"
%%% End: 

% LocalWords:  tuple reachability ZG iff Acc fract SG
 
%\input{inclusion} 
\section{Efficient testing of inclusion in a closure of a zone}
\label{sec:inclusion}

The tests of the form $Z\subseteq Closure_\a(Z')$ will be at the core
of the new algorithm we propose. This is an important difference with
respect to the standard algorithm that makes the tests of the form
$Z\incl Z'$. The latter tests are done in $\Oo(|X|^2)$ time, where
$|X|$ is the number of clocks. We present in this section a simple
algorithm that can do the tests $Z\subseteq Closure_\a(Z')$ at the
same complexity with neither the need to represent nor to compute the
closure.

We start by examining the question as to how one decides if a region $R$
intersects a zone $Z$. The important point is that it is enough to
verify that the projection on every pair of variables is
nonempty. This is the cornerstone for the efficient inclusion testing
algorithm that even extends to LU-approximations.

\subsection{When is $R\cap Z$ empty}

It will be very convenient to represent zones by \emph{distance
  graphs}. Such a graph has clocks as vertices, with an additional
special clock $x_0$ representing constant $0$. For readability, we
will often write $0$ instead of $x_0$. Between every two vertices
there is an edge with a weight of the form $(\fleq, c)$ where $c\in
\mathbb{Z}\cup\set{\infty}$ and $\fleq$ is either $\leq$ or $<$. An
edge $x\act{\fleq c} y$ represents a constraint $y-x\fleq c$: or in
words, the distance from $x$ to $y$ is bounded by $c$. Let
$\sem{G}$ be the set of valuations of clock variables satisfying all
the constraints given by the edges of $G$ with the restriction that
the value of $x_0$ is $0$.

An arithmetic over the weights $(\fleq, c)$ can be defined as
follows~\cite{bengtsson2004timed}.
\begin{description}
\item \emph{Equality} $(\fleq_1,c_1) = (\fleq_2,c_2)$ if $c_1 = c_2$
  and $\fleq_1 = \fleq_2$.
\item \emph{Addition} $(\fleq_1,c_1) + (\fleq_2,c_2) = (\fleq,c_1 +
  c_2)$ where $\fleq = <$ iff either $\fleq_1$ or $\fleq_2$ is $<$.
\item \emph{Minus} $-(\fleq,c) = (\fleq, -c)$. 
\item \emph{Order} $(\fleq_1,c_1) < (\fleq_2,c_2)$ if either $c_1 <
  c_2$ or ($c_1 = c_2$ and $\fleq_1 = < $ and $\fleq_2 =
  \le$). 
\item \emph{Floor} $\floor{(<,c)}=(\leq,c-1)$ and $\floor{(\leq,c)}=(\leq,c)$.
\end{description}
This arithmetic lets us talk about the weight of a path as a weight of
the sum of its edges. A cycle in a distance graph $G$ is said to be
\emph{negative} if the sum of the weights of its edges is at most
$(<,0)$; otherwise the cycle is \emph{positive}. The following useful
proposition is folklore.

\begin{proposition}\label{prop:cycles}
  A distance graph $G$ has only positive cycles iff $\sem{G}\not=\es$.
\end{proposition}

A distance graph is in \emph{canonical form} if the weight of the edge
from $x$ to $y$ is the lower bound of the weights of paths from $x$ to
$y$. A \emph{distance graph of a region $R$}, denoted $G_R$, is the
canonical graph representing all the constraints defining
$R$. Similarly $G_Z$ for a zone $Z$.

We can now state a necessary and sufficient condition for the
intersection $R\cap Z$
to be empty in terms of cycles in distance graphs. We denote by
$R_{xy}$ the weight of the edge $x \xra{\fleq_{xy} c_{xy}} y$ in the
canonical distance graph representing $R$. Similarly for $Z$.

\begin{proposition}\label{prop:intersection region zone}
  Let $R$ be a region and let $Z$ be a zone. The intersection $R \cap
  Z$ is empty iff there exist variables $x,y$ such that $Z_{yx} +
  R_{xy} \le (<,0)$.
\end{proposition}

A variant of this fact has been proven as an intermediate step of
Proposition~2 in~\cite{Bouyer:FMSD:2004}. 
%We give a different proof
%that allows us to derive an efficient inclusion test presented below.

\subsection{Efficient inclusion testing}

Our goal is to efficiently perform the test $Z\incl \Closure(Z')$ for
two zones $Z$ and $Z'$. We are aiming at $\Oo(|X|^2)$ complexity,
since this is the complexity of current algorithms used for checking
inclusion of two zones. Proposition~\ref{prop:intersection region
  zone} can be used to efficiently test the inclusion $R\incl
\Closure(Z')$. It remains to understand what are the regions
intersecting the zone $Z$ and then to consider all possible cases.
The next lemma basically says that every consistent instantiation of
an edge in $G_Z$ leads to a region intersecting $Z$.

\begin{lemma}\label{lem:values}
  Let $G$ be a distance graph in canonical form, with all cycles
  positive. Let $x, y$ be two variables, and let $x \act{\fleq_{xy}
    c_{xy}} y$ and $y \act{\fleq_{yx} c_{yx}} x$ be edges in $G$. For
  every $d \in \mathbb{R}$ such that $d \fleq_{xy} c_{xy} $ and $-d
  \fleq_{yx} c_{yx} $ there exists a valuation $v \in \sem{G}$
  with $v(y) - v(x) = d$.
\end{lemma}
Thanks to this lemma it is enough to look at edges of $G_Z$ one by one
to see what regions we can get. This insight is used to get the
desired efficient inclusion test

\begin{theorem} \label{thm:2clocks}
 Let $Z, Z'$ be zones. Then, $Z \nsubseteq \Closure_\a(Z')$ iff there
 exist variables $x$, $y$, both different from $x_0$, such that one of
 the following conditions hold:
 \begin{enumerate}
   \item $Z'_{0x} < Z_{0x}$ and $Z'_{0x} \le (\le,\a_x)$, or
   \item $Z'_{x0} < Z_{x0}$ and $Z_{x0} \ge (\le,-\a_x)$, or 
   \item $Z_{x0} \ge (\le,-\a_x)$ and $Z'_{xy} < Z_{xy}$ and
     $Z'_{xy} \le (\le,\a_y) + \floor{Z_{x0}}  $.  
    \end{enumerate}
\end{theorem}

\subsubsection*{Comparison with the algorithm for $Z \subseteq Z'$}
Given two zones $Z$ and $Z'$, the procedure for checking $Z \subseteq
Z'$ works on two graphs $G_Z$ and $G_{Z'}$ that are in canonical
form. This form reduces the inclusion test to comparing the edges of
the graphs one by one. Note that our algorithm for $Z \subseteq
Closure_\a(Z')$ does not do worse. It works on $G_Z$ and $G_{Z'}$
too. The edge by edge checks are only marginally more complicated. The
overall procedure is still $\Oo(|X|^2)$.

\subsection{Handling LU-approximation}
\label{sec:LU-effcient}

In~\cite{Behrmann:STTT:2006} the authors propose to distinguish
between maximal constants used in upper and lower bounds comparisons:
for each clock $x$, $L_x \in \Nat \cup \{ -\infty\}$ represents the
maximal constant $c$ such that there exists a constraint $x > c$ or $x
\ge c$ in a guard of a transition in the automaton; dually, $U_x \in
\Nat \cup \{-\infty\}$ represents the maximal constant $c$ such that
there is a constraint $x < c$ or $x \le c$ in a guard of a
transition. If such a $c$ does not exist, then it is considered to be
$-\infty$.  They have introduced an extrapolation operator
$\Extra^+_{LU}(Z)$ that takes into account this information. This is
probably the best presently known convex abstraction of zones.

We now explain how to extend our inclusion test to handle LU
approximation, namely given $Z$ and $Z'$ how to directly check $Z
\incl \Closure_\a(\Extra^+_{LU}(Z'))$ efficiently. Observe that for
each $x$, the maximal constant $\a_x$ is the maximum of $L_x$ and
$U_x$.  In the sequel, this is denoted $Z \incl
\Closure_{LU}^+(Z')$. For this we need to understand first when a
region intersecting $Z$ intersects $\Extra^+_{LU}(Z')$. Therefore, we
study the conditions that a region $R$ should satisfy if it intersects
$\Extra^+_{LU}(Z)$ for a zone $Z$.

We recall the definition given in \cite{Behrmann:STTT:2006} that has
originally been presented using difference bound matrices (DBM). In a
DBM $(c_{ij}, \prec_{i,j})$ stands for $x_i - x_j \prec_{i,j}
c_{i,j}$. In the language of distance graphs, this corresponds to an
edge $x_j \act{\prec_{i,j} c_{i,j}} x_i$; hence to $Z_{ji}$ in our
notation.  Let $Z^+$ denote $\Extra^+_{LU}(Z)$ and $G_{Z^+}$ its
distance graph. We have:
{\small
\begin{equation}\label{lu-defn}
  Z^+_{xy} = \begin{cases}
    (<,\infty) & \text{if } Z_{xy} > (\le, L_y) \\
    (<,\infty) & \text{if }  -Z_{y0} > (\le, L_y)\\
    (<,\infty) & \text{if } -Z_{x0} > (\le,U_x), y \neq 0 \\
    (<,-U_x) & \text{if } -Z_{x0} > (\le,U_x), y = 0 \\
    Z_{xy} & \text{otherwise}.
  \end{cases}
\end{equation}
}
From this definition it will be important for us to note that
$G_{Z^+}$ is $G_Z$ with some weights put to $(<,\infty)$ and some
weights on the edges to $x_0$ put to $(<,-U_x)$.  Note that
$\Extra^+_{LU}(Z')$ is not in the canonical form.  If we put
$\Extra^+_{LU}(Z')$ into the canonical form then we could just use
Theorem~\ref{thm:2clocks}. We cannot afford to do this since
canonization can take cubic time \cite{bengtsson2004timed}.  The
following theorem implies that we can do the test without
canonizing $\Extra^+_{LU}(Z')$. Hence we can get a simple quadratic test
also in this case.

\begin{theorem} 
  Let $Z, Z'$ be zones. Let $Z'^+$ denote $\Extra^+_{LU}(Z')$ obtained
  from $Z'$ using Equation \ref{lu-defn} for each edge. Note that
  $Z'^+$ is not necessarily in canonical form. Then, we get that $Z
  \nsubseteq \Closure_\a(Z'^+)$ iff there exist variables $x$, $y$
  different form $x_0$ such that one of the following conditions hold:
  \begin{enumerate}
  \item $Z'^+_{0x} < Z_{0x}$ and $Z'^+_{0x} \le (\le,\a_x)$, or
  \item $Z'^+_{x0} < Z_{x0}$ and $Z_{x0} \ge (\le,-\a_x)$, or
  \item $Z_{x0} \ge (\le, -\a_x)$ and $Z'^+_{xy} < Z_{xy}$ and
    $Z'^+_{xy} \le (\le,\a_y) + \floor{Z_{x0}} $.
  \end{enumerate}
\end{theorem}

%%% Local Variables: 
%%% mode: latex
%%% TeX-master: "m"
%%% End: 

% LocalWords: igw iff xy wrt DBM ij

\section{A New Algorithm for Reachability}
\label{sec:lazy}

Our goal is to decide if a final state of a given timed automaton is
reachable.  We do it by computing a finite prefix of the
reachability tree of the zone graph $ZG(\Aa)$ that is sufficient to
solve the reachability problem. Finiteness is ensured by not exploring
a node $(q,Z)$ if there exists a $(q,Z')$ such that $Z \subseteq
Closure_{\a}(Z')$, for a suitable $\a$. We will first describe a
simple algorithm based on the closure and then we will address the
issue of finding tighter bounds for the clock values.

\subsection{The basic algorithm}

Given a timed automaton $\Aa$ we first calculate the bound function
$\a_\Aa$ as described just before
Theorem~\ref{thm:symbolic-zone-graph}. Each node in the tree that we
compute is of the form $(q,Z)$, where $q$ is a state of the automaton,
and $Z$ is an unapproximated zone. The root node is $(q_0,Z_0)$, which
is the initial node of $ZG(\Aa)$. The algorithm performs a depth first
search: at a node $(q,Z)$, a transition $t = (q,g,r,q')$ not yet
considered for exploration is picked and the successor $(q',Z')$ is
computed where $(q,Z) \xrightarrow{t} (q',Z')$ in $ZG(\Aa)$. If $q'$
is a final state and $Z'$ is not empty then the algorithm
terminates. Otherwise the search continues from $(q',Z')$ unless there
is already a node $(q',Z'')$ with $Z'\incl Closure_{\a_\Aa}(Z'')$ in
the current tree.

The correctness of the algorithm is straightforward. It follows from
the fact that if $Z'\incl \Closure_{\a_\Aa}(Z'')$ then all the states
reachable from $(q',Z')$ are reachable from $(q',Z'')$ and hence it is
not necessary to explore the tree from $(q',Z')$. Termination of the
algorithm is ensured since there are finitely many sets of the form
$\Closure_{\a_\Aa}(Z)$. Indeed, the algorithm will construct a prefix
of the reachability tree of $SG_{\a}(\Aa)$ as described in
Theorem~\ref{thm:symbolic-zone-graph}.

The above algorithm does not use the classical extrapolation
operator named $Extra_M^+$ in~\cite{Behrmann:STTT:2006} and
$Extra_\a^+$ hereafter, but the coarser $\Closure_\a$
operator~\cite{Bouyer:FMSD:2004}. This is possible since the algorithm
does not need to represent $\Closure_\a(Z)$, which is in general not a
zone. Instead of storing $\Closure_\a(Z)$ the algorithm just stores
$Z$ and performs tests $Z\incl \Closure_\a(Z')$ each time it is needed
(in contrast to Algorithm~2 in~\cite{Bouyer:FMSD:2004}). This is as
efficient as testing $Z\incl Z'$ thanks to the algorithm
presented in the previous section.

Since $\Closure_\a$ is a coarser abstraction, this simple algorithm
already covers some of the optimizations of the standard algorithm.
For example the $Extra^+_\a(Z)$ abstraction proposed
in~\cite{Behrmann:STTT:2006} is subsumed since $Extra^+_\a(Z)\incl
\Closure_\a(Z)$ for any zone $Z$ 
\cite{Bouyer:FMSD:2004,Behrmann:STTT:2006}. 
Other important optimizations of the
standard algorithm concern finer computation of bounding functions
$\a$. We now show that the structure of the proposed algorithm allows
to improve this too.

{
  \setlength{\columnsep}{30pt}
  \setlength{\columnseprule}{0.5pt}
\begin{lstlisting}[mathescape=true,basicstyle=\scriptsize,multicols=2,
columns=fullflexible,float=t,
caption={Reachability algorithm with on-the-fly bound computation and
non-convex abstraction.},label=fig:algorithm]
function main():
  push($(q_0,Z_0,\a_0)$, stack)
  while (stack $\neq$ $\es$) do
    $(q,Z,\a)$ := top(stack); pop(stack)
    explore($q,Z,\a$)
    resolve()
  return "empty"

function explore($q,Z,\a$):
  if ($q$ is accepting)
    exit "not empty"
  if ($\exists$ $(q,Z',\a')$ nontentative
            and s.t. $Z\subseteq\Closure_{\a'}(Z')$)
    mark $(q,Z,\a)$ tentative wrt $(q,Z',\a')$
    $\a$ := $\a'$; propagate($parent(q,Z,\a)$)
  else
    propagate($q,Z,\a$)
    for each $(q_s,Z_s,\a_s)$ in $children(q,Z,\a)$ do
      if ($Z_s$ $\neq$ $\es$)
        explore($q_s,Z_s,\a_s$)

function resolve():
  for each $(q,Z,\a)$ tentative wrt $(q,Z',\a')$ do
    if ($Z\not\subseteq\Closure_{\a'}(Z')$)
      mark $(q,Z,\a)$ nontentative
      $\a$ := $-\infty$; propagate($parent(q,Z,\a)$)
      push($(q,Z,\a)$, stack)

function propagate($q,Z,\a$):
  $\a$ := $max_{(q,Z,\a)\xra{g;R}(q',Z',\a')}$ maxedge($g,R,\a'$)
  if ($\a$ has changed)
    for each $(q_t,Z_t,\a_t)$ tentative wrt $(q,Z,\a)$ do
      $\a_t$ := $\a$; propagate($parent(q_t,Z_t,\a_t)$)
    if ($(q,Z,\a)\neq(q_0,Z_0,\a_0)$)
      propagate($parent(q,Z,\a)$)

function maxedge($g,R,\a$):
  let $\a_R= \lambda x.$ if $x\in R$ then $-\infty$ else $\a(x)$
  let $\a_g= \lambda x.$ if $x\#c$ in $g$ then $c$ else $-\infty$
  return ($\lambda x.$ $max(\a_R(x),\a_g(x))$)
\end{lstlisting}
}

\subsection{Computing clock bounds on-the-fly}
\label{sec:bounds}

We can improve on the idea of Behrmann et
al.~\cite{Behrmann:TACAS:2003} of computing a bound function $\a_q$ for
each state $q$. We will compute these bounding functions on-the-fly
and they will depend also on a zone and not just a state.  An obvious
gain is that we will never consider constraints coming from
unreachable transitions. We comment more on advantages of this
approach in Section~\ref{sec:discussion}.

Our modified algorithm is given in Figure~\ref{fig:algorithm}. It
computes a tree whose nodes are triples $(q,Z,\a)$ where $(q,Z)$ is a
node of $ZG(\Aa)$ and $\a$ is a bound function. Each node $(q,Z,\a)$
has as many child nodes $(q_s,Z_s,\a_s)$ as there are successors
$(q_s,Z_s)$ of $(q,Z)$ in $ZG(\Aa)$. Notice that this includes
successors with an empty zone $Z_s$, which are however not further
unfolded. These nodes must be included for correctness of our constant
propagation procedure. By default bound functions map each clock to
$-\infty$. They are later updated as explained below. Each node is
further marked either $tentative$ or $nontentative$. The leaf nodes
$(q,Z,\a)$ of the tree are either deadlock nodes (either there is no
transition out of state $q$ or $Z$ is empty), or $tentative$
nodes. All the other nodes are marked $nontentative$.

Our algorithm starts from the root node $(q_0,Z_0,\a_0)$, consisting
of the initial state, initial zone, and the function mapping each clock
to $-\infty$.  It repeatedly alternates an exploration and a
resolution phase as described below.

\subsubsection*{Exploration phase}

Before exploring a node $n=(q,Z,\a)$ the function \verb|explore|
checks if $q$ is accepting and $Z$ is not empty; if it is so then
$\Aa$ has an accepting run. Otherwise the algorithm checks if there
exists a $nontentative$ node $n'=(q',Z',\a')$ in the current tree such
that $q=q'$ and $Z\subseteq\Closure_{\a'}(Z')$. If yes, $n$ becomes a
$tentative$ node and its exploration is temporarily stopped as each
state reachable from $n$ is also reachable from $n'$. If none of these
holds, the successors of the node are explored. The exploration
terminates since $\Closure_\a$ has a finite range. 

When the exploration algorithm gets to a new node, it propagates the
bounds from this node to all its predecessors. The goal of these
propagations is to maintain the following invariant. For every node
$n=(q,Z,\a)$:
\begin{enumerate}
\item if $n$ is $nontentative$, then $\a$ is the maximum of the $\a_s$
  from all successor nodes $(q_s,Z_s,\a_s)$ of $n$ (taking into account
  guards and resets as made precise in the function \verb|maxedge|);
\item if $n$ is $tentative$ with respect to $(q',Z',\a')$, then $\a$
  is equal to $\a'$.
\end{enumerate}
The result of propagation is analogous to the inequalities seen in the
static guard analysis~\cite{Behrmann:TACAS:2003}, however now applied
to the zone graph, on-the-fly. Hence, the bounds associated to each
node $(q,Z,\a)$ never exceed those that are computed by the static
guard analysis.

A delicate point about this procedure is handling of tentative
nodes. When a node $n$ is marked $tentative$, we have
$\a=\a'$. However the value of $\a'$ may be updated when the tree is
further explored. Thus each time we update the bounds function of a
node, it is not only propagated upward in the tree but also to the
nodes that are tentative with respect to $n'$.

This algorithm terminates as the bound functions in each node never
decrease and are bounded. From the invariants above, we get that
in every node, $\a$ is a solution to the equations
in~\cite{Behrmann:TACAS:2003} applied on $ZG(\Aa)$.

It could seem that the algorithm will be forced to do a high number of
propagations of bounds. The experiments reported in
Section~\ref{sec:discussion} show that the present very simple
approach to bound propagation is good enough. Since we propagate the
bounds as soon as they are modified, most of the time, the value of
$\a$ does not change in line~30 of function \verb|propagate|. In
general, bounds are only propagated on very short distances in the
tree, mostly along one single edge. For this reason we do not
concentrate on optimizing the function \verb|propagate|. In the
implementation we use the presented function augmented with a minor
``optimization'' that avoids calculating maximum over all successors
in line~30 when it is not needed.

\subsubsection*{Resolution phase}

Finally, as the bounds may have changed since $n$ has been marked
tentative, the function \verb|resolve| checks for the consistency of
$tentative$ nodes. If $Z\subseteq\Closure_{\a'}(Z')$ is not true
anymore, $n$ needs to be explored. Hence it is viewed as a new node:
the bounds are set to $-\infty$ and $n$ is pushed on the $stack$ for
further consideration in the function \verb|main|. Setting $\a$ to
$-\infty$ is safe as $\a$ will be computed and propagated when $n$ is
explored.  We perform also a small optimization and propagate this
bound upward, thereby making some bounds decrease.

The resolution phase may provide new nodes to be explored. The
algorithm terminates when this is not the case, that is when all
tentative nodes remain tentative. We can then conclude that no
accepting state is reachable.

\begin{theorem}\label{thm:alg correct}
  An accepting state is reachable in $ZG(\Aa)$ iff the
  algorithm reaches a node with an accepting state and a non-empty zone. 
\end{theorem}

\subsection{Handling LU approximations}

Recall that $\Extra^+_{LU}(Z)$ approximation used two bounds: $L_x$
and $U_x$ for each clock $x$.  In our algorithm we can easily
propagate LU bounds instead of just maximal bounds. We can also
replace the test $Z\incl \Closure_{\a'}(Z')$ by $Z\incl
\Closure_{\a'}(\Extra^+_{L'U'}(Z'))$, where $L'$ and $U'$ are the
bounds calculated for $(q',Z')$ and $\a'_x=\max(L'_x,U'_x)$ for every
clock $x$. As discussed in Section~\ref{sec:LU-effcient}, this test
can be done efficiently too. The proof of correctness of the resulting
algorithm is only slightly more complicated.

%%% Local Variables: 
%%% mode: latex
%%% TeX-master: "m"
%%% End: 

% LocalWords: Reachability reachability ZG nontentative todo
% nontentative Acc LocalWords: iff
 
\begin{table}[t]
  \scriptsize
  \begin{center}
    \begin{tabular}{|c||r|r||r|r||r|r|}
      \hline
      Model
      & \multicolumn{2}{|c||}{Our algorithm}
      & \multicolumn{2}{|c||}{UPPAAL's algorithm}
      & \multicolumn{2}{|c|}{UPPAAL 4.1.3 (-n4 -C -o1)}
      \\
      \cline{2-7}
      & nodes & s.
      & nodes & s.
      & nodes & s.
      \\
      \hline
      $\Aa_1$
      & $2$ & $0.00$
      & $10003$ & $0.07$
      & $10003$ & $0.07$
      \\
      $\Aa_2$
      & $7$ & $0.00$
      & $3999$ & $0.60$
      & $2003$ & $0.01$
      \\
      $\Aa_3$
      & $3$ & $0.00$
      & $10004$ & $0.37$
      & $10004$ & $0.32$
      \\
      \hline
      CSMA/CD7
      & $5031$ & $0.32$
      & $5923$ & $0.27$
      & $-$ & T.O.
      \\
      CSMA/CD8
      & $16588$ & $1.36$
      & $19017$ & $1.08$
      & $-$ & T.O.
      \\
      CSMA/CD9
      & $54439$ & $6.01$
      & $60783$ & $4.19$
      & $-$ & T.O.
      \\
      \hline
      FDDI10
      & $459$ & $0.02$
      & $525$ & $0.06$
      & $12049$ & $2.43$
      \\
      FDDI20
      & $1719$ & $0.29$
      & $2045$ & $0.78$
      & $-$ & T.O.
      \\
      FDDI30
      & $3779$ & $1.29$
      & $4565$ & $4.50$
      & $-$ & T.O.
      \\
      \hline
      Fischer7
      & $7737$ & $0.42$
      & $20021$ & $0.53$
      & $18374$ & $0.35$
      \\
      Fischer8
      & $25080$ & $1.55$
      & $91506$ & $2.48$
      & $85438$ & $1.53$
      \\
      Fischer9
      & $81035$ & $5.90$
      & $420627$ & $12.54$
      & $398685$ & $8.95$
      \\
      Fischer10
      & $-$ & T.O.
      & $-$ & T.O.
      & $1827009$ & $53.44$
      \\
      \hline
    \end{tabular}
  \end{center}
  \caption{Experimental results: number of visited nodes and running
    time with a timeout (T.O.) of 60 seconds. Experiments done
    on a MacBook with 2.4GHz Intel Core Duo processor and 2GB of
    memory running MacOS X 10.6.7.}
  \label{fig:experimental-results}
\end{table}

\section{Experimental results}
\label{sec:discussion}

We have implemented the algorithm from Figure~\ref{fig:algorithm}, and
have tested it on classical benchmarks. The results are presented in
Table~\ref{fig:experimental-results}, along with a comparison to
UPPAAL and our implementation of UPPAAL's core algorithm that uses the
$Extra_{LU}^+$ extrapolation~\cite{Behrmann:STTT:2006} and computes
bounds by static analysis~\cite{Behrmann:TACAS:2003}. Since we have
not considered symmetry reduction~\cite{Hendriks:FORMATS:2003} in our
tool, we have not used it in UPPAAL either.

The comparison to UPPAAL is not meaningful for the CSMA/CD and the
FDDI protocols.  Indeed, UPPAAL runs out of time even if we
significantly increase the time allowed; switching to breadth-first
search has not helped either.  We suspect that this is due to the
order in which UPPAAL takes the transitions in the automaton. For this
reason in columns~4 and~5, we provide results from our own
implementation of UPPAAL's algorithm that takes transitions in the
same order as the implementation of our algorithm. Although RED also
uses approximations, it is even more difficult to draw a meaningful
comparison with it, since it uses symbolic state representation unlike
UPPAAL or our tool. Since this paper is about approximation methods,
and not tool comparison, we leave more extensive comparisons as
further work.

The results show that our algorithm provides important
gains. Analyzing the results more closely we could see that both the
use of closure, and on-the-fly computation of bounds are important. In
Fischer's protocol our algorithm visits much less nodes. In the FDDI
protocol with $n$ processes, the DBMs are rather big square matrices
of order $3n+2$. Nevertheless our inclusion test based on $Closure$ is
significantly better in the running time. The CSMA/CD case shows that
the cost of bounds propagation does not always counterbalance the
gains. However the overhead is not very high either. We comment
further on the results below.

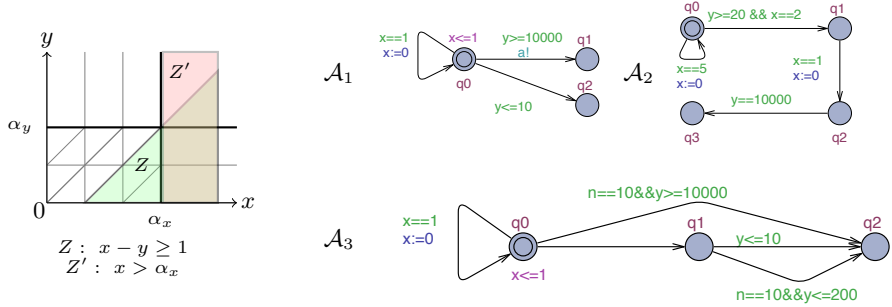
\begin{figure}[t]
  \begin{minipage}{.3\textwidth}
    \begin{tikzpicture}[scale=0.5]
  % Regions
  \foreach \x in {0,...,3} \draw[very thin, gray] (\x,0) -- (\x,4);
  \foreach \x in {1, ..., 3} \draw[very thin, gray] (\x,0) -- (3, 3
  - \x); \draw[very thin, gray] (0,0) -- (2,2); \foreach \y in
  {0,...,2} { \draw[very thin, gray] (0,\y) -- (5,\y); \draw[very
    thin, gray] (0,\y) -- (2 - \y, 2); }
    
  % bounds
  \draw[thick] (3,0) -- (3,4);
  \draw[thick] (0,2) -- (5,2);

  % Axes
  \draw[->] (0,0) -- (0,4); \draw[->] (0,0) -- (5,0);
  
  % Z
  \draw[thick,fill=green!50,nearly transparent] (1,0) -- (4.5,3.5) --
  (4.5,0) -- cycle;
  \draw (2.5,1) node {\scriptsize $Z$};

  % Z'
  \draw[thick,fill=red!50,nearly transparent] (3.05,0) -- (3.05,4) --
  (4.5,4) -- (4.5,0) -- cycle;
  \draw (3.5,3.5) node {\scriptsize $Z'$};

  % Naming
  \draw (3,-0.5) node {\scriptsize{$\a_x$}} (4,-0.5);
  \draw (-0.7, 2) node {\scriptsize{$\a_y$}} (0,2); 
  \draw (5.3,0) node {$x$} (6, 0);
  \draw (0, 4.3) node {$y$} (0,4.3);
  \draw (-0.2,-0.2) node {$0$} (0,-0.2);

  \draw (2,-1.2) node {\scriptsize $Z:\ x-y\geq 1$};
  \draw (2,-1.7) node {\scriptsize $Z':\ x>\a_x$};
\end{tikzpicture}

%%% Local Variables: 
%%% mode: latex
%%% TeX-master: "../m"
%%% End: 
  \end{minipage}
  \hfill
  \begin{minipage}{.65\textwidth}
    \begin{minipage}{.07\textwidth}
      $\Aa_1$
    \end{minipage}
    \begin{minipage}{.4\textwidth}
      \includegraphics[scale=0.15]{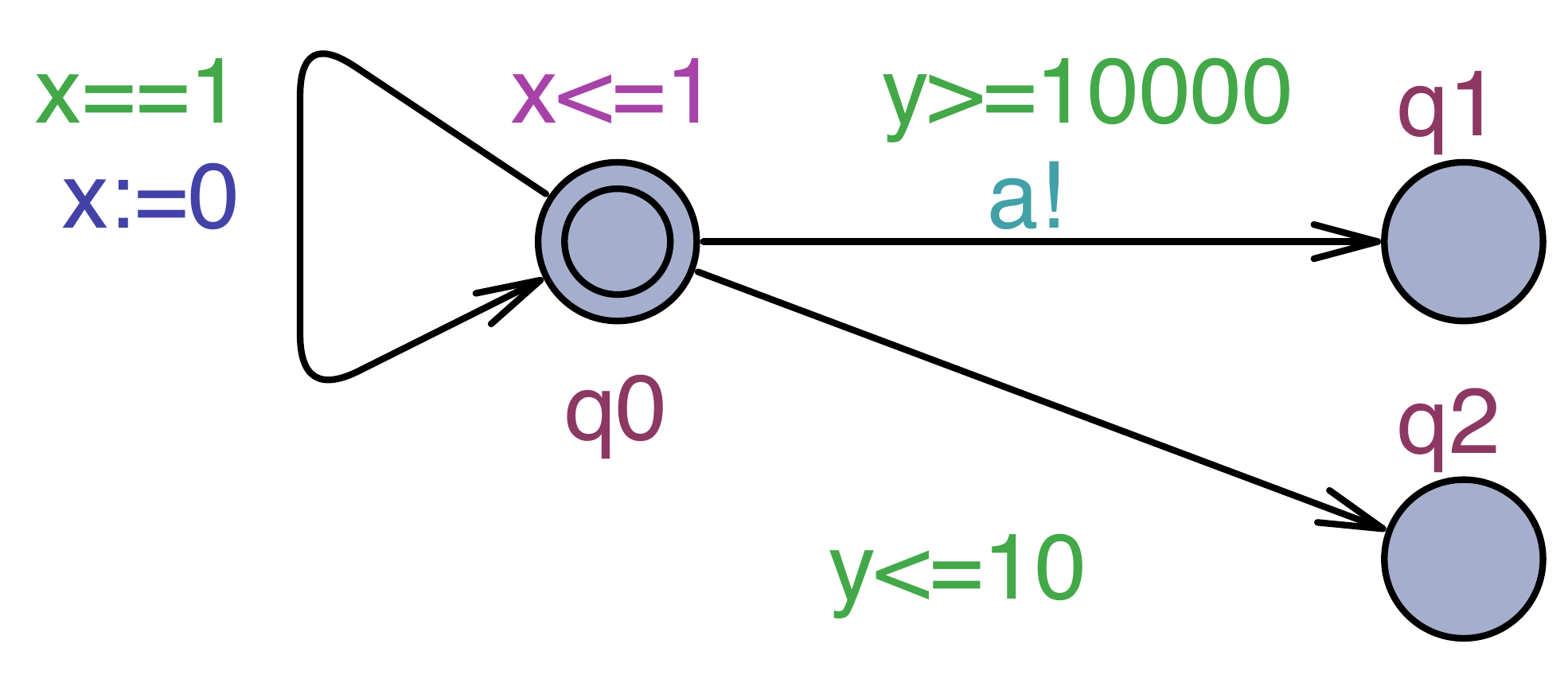}
    \end{minipage}
    \begin{minipage}{.07\textwidth}
      $\Aa_2$
    \end{minipage}
    \begin{minipage}{.4\textwidth}
      \includegraphics[scale=0.12]{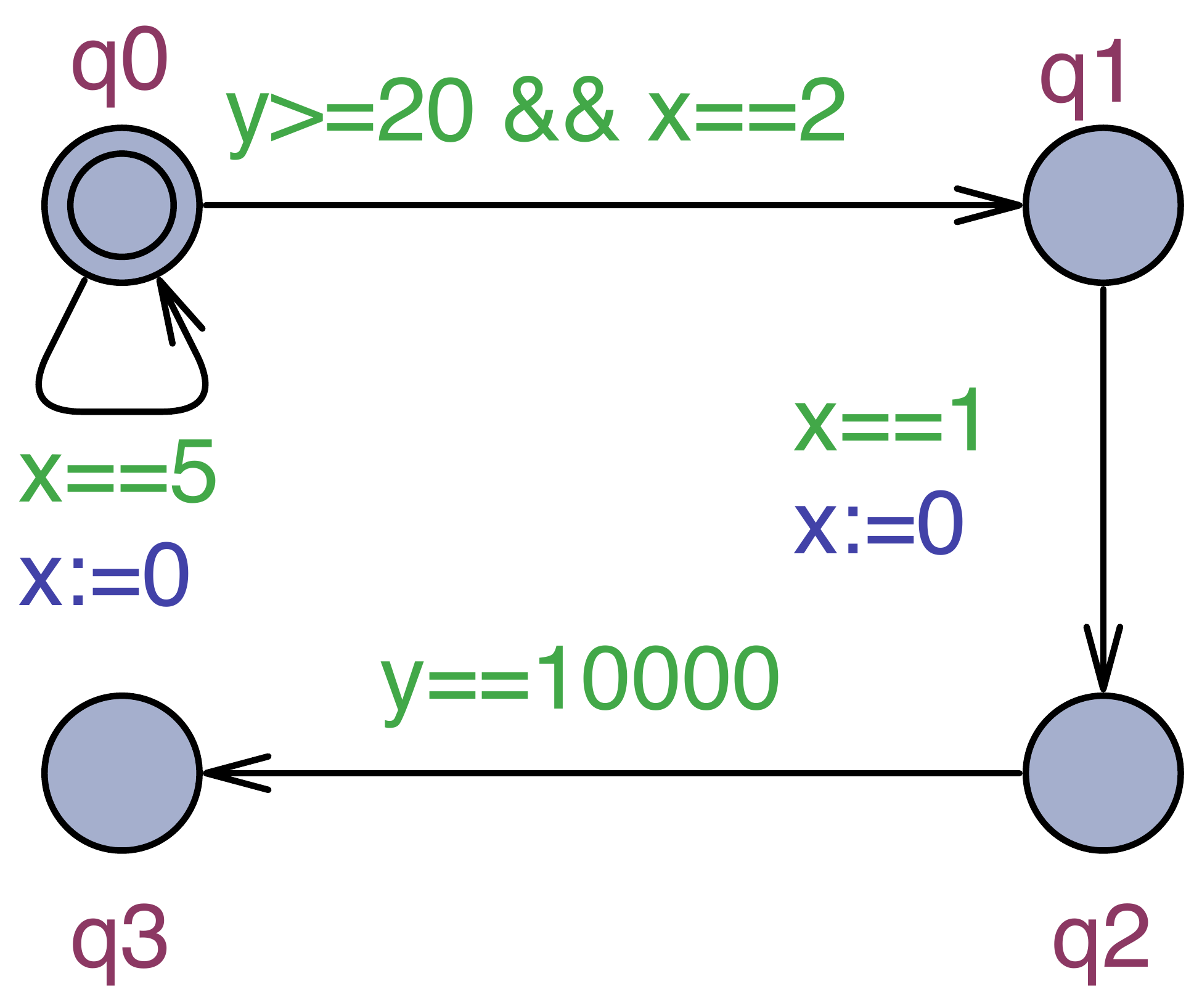}
    \end{minipage}
    
    \medskip
    \begin{minipage}{.1\textwidth}
      $\Aa_3$
    \end{minipage}
    \begin{minipage}{.85\textwidth}
      \includegraphics[scale=0.33]{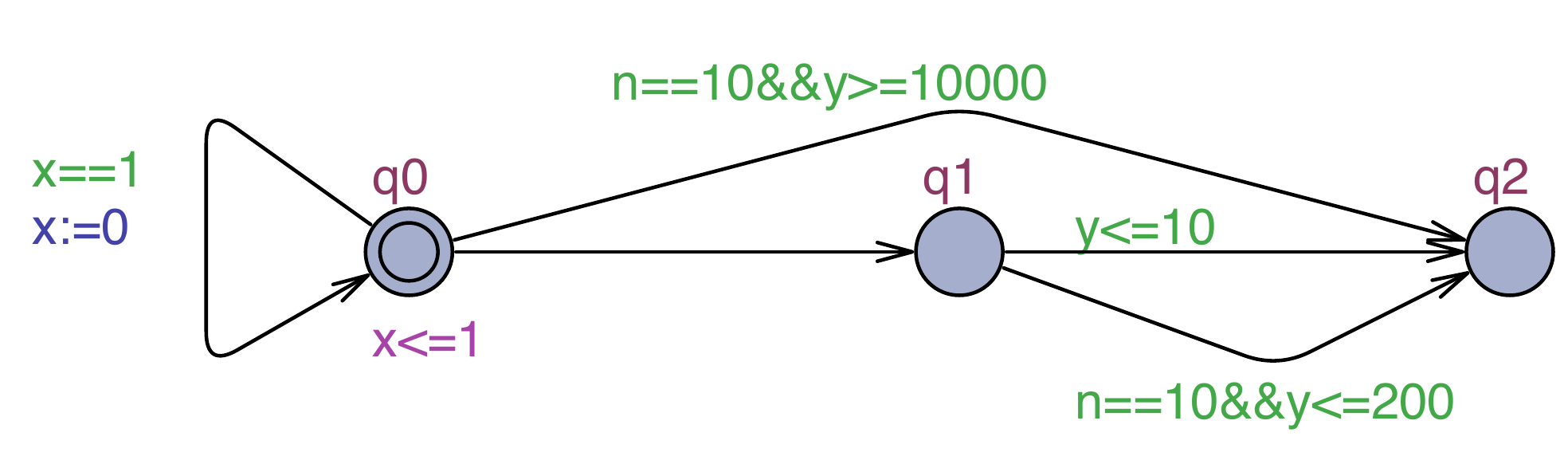}
    \end{minipage}
  \end{minipage}

  \caption{Examples explaining gains obtained with the algorithm.}
  \label{fig:uppaal}
\end{figure}

The first improvement comes from the computation of the maximal bounds
used for the abstraction as demonstrated by the examples $\Aa_2$
(Figure~\ref{fig:uppaal}), Fischer and CSMA/CD that correspond to
three different situations. In the $\Aa_2$ example, the transition that
yields the big bound $10^4$ on $y$ in $q_0$ is not reachable from any
$(q_0,Z)$, hence we just get the lower bound $20$ on $y$ in $(q_0,Z)$, and
a subsequent gain in performance.

The automaton $\Aa_1$ in Figure~\ref{fig:uppaal} illustrates the gain
on the CSMA/CD protocol. The transition from $q_0$ to $q_1$ is
disabled as it must synchronize on letter $a!$. The static analysis
algorithm~\cite{Behrmann:TACAS:2003} ignores this fact, hence it
associates bound $10^4$ to $y$ in $q_0$. Since our algorithm computes
the bounds on-the-fly, $y$ is associated the bound $10$ in every node
$(q_0,Z)$. We observe that UPPAAL's algorithm visits $10003$ nodes on
$ZG(\Aa_1)$ whereas our algorithm only visits $2$ nodes. The same
situation occurs in the CSMA/CD example. However despite the
improvement in the number of nodes (roughly $10\%$) the cost of
computing the bounds impacts the running time
negatively.

The gains that we observe in the analysis of the Fischer's protocol
are explained by the automaton $\Aa_3$ in
Figure~\ref{fig:uppaal}. $\Aa_3$ has a bounded integer variable $n$
that is initialized to $0$. Hence, the transitions from $q_0$ to
$q_2$, and from $q_1$ to $q_2$, that check if $n$ is equal to $10$ are
disabled. This is ignored by the static analysis algorithm that
associates the bound $10^4$ to clock $y$ in $q_0$. Our algorithm
however associates the bound $10$ to $y$ in every node $(q_0,Z)$. We
observe that UPPAAL's algorithm visits $10004$ nodes whereas our
algorithm only visits $3$ nodes. A similar situation occurs in the
Fischer's protocol. We include the last row to underline that our
implementation is not as mature as UPPAAL. We strongly think that
UPPAAL could benefit from methods presented here.

The second kind of improvement comes from the $\Closure_\a$
abstraction that particularly improves the analysis of the Fischer's
and the FDDI protocols. The situation observed on the FDDI protocol is
explained in Figure~\ref{fig:uppaal}.  For the zone $Z$ in the figure,
by definition $Extra_{LU}^+(Z)=Z$, and in consequence $Z'\not\subseteq
Z$. However, $Z'\subseteq \Closure_\a(Z)$. On FDDI and Fischer's
protocols, our algorithm performs better due to the non-convex
approximation.

\section{Conclusions}

We have proposed a new algorithm for checking reachability properties
of timed automata. The algorithm has two sources of improvement that
are quite independent: the use of the $\Closure_\a$ operator, and the
computation of bound functions on-the-fly.

Apart from immediate gains presented in
Table~\ref{fig:experimental-results}, we think that our approach opens
some new perspectives on analysis of timed systems. We show that the
use of non-convex approximations can be efficient. We have used very
simple approximations, but it may be well the case that there are more
sophisticated approximations to be discovered. The structure of our
algorithm permits to calculate bounding constants on the fly. One
should note that standard benchmarks are very well understood and very
well modeled. In particular they have no ``superfluous'' constraints
or clocks. However in not-so-clean models coming from systems in
practice one can expect the on-the-fly approach to be even more
beneficial.

There are numerous directions for further research. One of them is to
find other approximation operators. Methods for constraint propagation
also deserve a closer look. We believe that our approximations methods
are compatible with partial order
reductions~\cite{Hendriks:FORMATS:2003,Malinowski:TACAS:2010}. We hope
that the two techniques can benefit from each other.

%%% Local Variables: 
%%% mode: latex
%%% TeX-master: "m"
%%% End: 

% Bibliography
\bibliographystyle{plain} \bibliography{m}
\appendix
\section{Proofs from Section 3}
We provide all the proofs from the section presenting the efficient
inclusion testing algorithm. For convenience, we recall the statements of the
facts that are proven together with their original
numbering. They are preceded with black arrow for readability.

\Repeat{Proposition}{prop:cycles} A distance graph $G$ has only
positive cycles iff $\sem{G}\not=\es$.

\begin{proof}
  If there is a valuation $v\in\sem{G}$ then we replace every edge
  $x\act{\fleq c} y$ by $x\act{\leq d} y$ where $d=v(y)-v(x)$. We have
  $d\fleq_{xy} c_{xy}$. Since every cycle in the new graph has value
  $0$, every cycle in $G$ is positive.

  For the other direction suppose that every cycle in $G$ is
  positive. Let $\bar G$ be the canonical form of $G$. Clearly
  $\sem{G}=\sem{\bar G}$, i.e., the constraints defined by $G$ and by
  $\bar G$ are equivalent. It is also evident that all the cycles in
  $\bar G$ are positive.

  We say that a variable $x$ is \emph{fixed} in $\bar G$ if in this
  graph we have edges $0\act{\leq c_x} x$ and $x\act{\leq -c_x} 0$ for
  some constant $c_x$. These edges mean that every valuation in
  $\sem{\bar G}$ should assign $c_x$ to $x$.

  If all the variables in $\bar G$ are fixed then the value of every
  cycle in $\bar G$ is $0$, and the valuation assigning $c_x$ to $x$
  for every variable $x$ is the unique valuation in $\sem{\bar
    G}$. Hence, $\sem{\bar G}$, and in consequence $\sem{G}$ are not
  empty.

  Otherwise there is a variable, say $y$, that is not fixed in $\bar
  G$. We will show how to fix it. Let us multiply all the constraints
  in $\bar G$ by $2$. This means that we change each arrow
  $x_1\act{\fleq c} x_2$ to $x_1\act{\fleq 2c} x_2$. Let us call the
  resulting graph $H$. Clearly $H$ is in canonical form since $\bar
  G$ is. Moreover $\sem{H}$ is not empty iff $\sem{\bar G}$ is not
  empty. The gain of this transformation is that for our chosen
  variable $y$ we have in $H$ edges $0\act{\fleq c_{0y}} y$ and
  $y\act{\fleq c_{y0}} 0$ with $c_{y0}+c_{0y}\geq 2$. This means that
  there is a natural number $d$ such that $(\leq,d)\leq (\fleq
  c_{0y})$ and $(\leq,-d)\leq (\fleq c_{y0})$. Let $H_d$ be $H$ with
  edges to and from $y$ changed to $0\act{\leq d} y$ and $y\act{\leq
    -d} 0$, respectively. This is a distance graph where $y$ is
  fixed. We need to show that there is no negative cycle in this
  graph.

  Suppose that there is a negative cycle in $H_d$. Clearly it has to
  pass through $0$ and $y$ since there was no negative cycle in $H$. Suppose
  that it uses the edge $0\act{\leq d} y$, and suppose that the next
  used edge is $y\act{\fleq c_{yx}} x$. The cycle cannot come back to
  $y$ before ending in $0$ since then we could construct a smaller
  negative cycle. Hence all the other edges in the cycle come from
  $H$. Since $H$ is in the canonical form, a path from $x$ to $0$ can be
  replaced by the edge from $x$ to $0$, and the value of the path will
  not increase. This means that our hypothetical negative cycle has
  the form $0\act{\leq d} y\act{\fleq c_{yx}} x\act{\fleq
    c_{x0}}0$. By canonicity of $H$ we have
  $(\fleq_{yx},c_{yx})+(\fleq_{x0},c_{x0}) \geq
  (\fleq_{y0},c_{y0})$. Putting these two facts together we get
  \begin{equation*}
    (\leq,0)>(\leq,d)+(\fleq_{yx},c_{yx})+(\fleq_{x0},c_{x0})\geq
    (\leq, d)+(\fleq_{y0},c_{y0})
  \end{equation*}
  but this contradicts the choice of $d$ which supposed that $(\leq,
  d)+(\fleq_{y0},c_{y0})$ is positive. The proof when
  the hypothetical negative cycle passes through the edge $y\act{\leq
    -d} 0$ is analogous.

  Summarizing, starting from $G$ that has no negative cycles we have
  constructed a graph $H_d$ that has no negative cycles, and has one
  more variable fixed. We also know that if $\sem{H_d}$ is not empty
  then $\sem{G}$ is not empty. Repeatedly applying this construction
  we get a graph where all the variables are fixed and no cycle is
  negative. As we have seen above the semantics of such a graph is not
  empty.  \qed
\end{proof}

\Repeat{Proposition}{prop:intersection region zone} Let $R$ be a
region and let $Z$ be a zone. The intersection $R\cap Z$ is empty iff
there exist variables $x,y$ such that $Z_{yx} + R_{xy} \le (<,0)$.
\medskip

Before proving the above proposition, we will start with some notions.
Let $R$ be a region wrt. a bound function $\a:X\to \Nat$. A variable
$x$ is \emph{bounded in $R$} if a constraint $x\leq c$ holds in $R$
for some constant $c$; otherwise the variable is called
\emph{unbounded in $R$}. Observe that if $x_1,x_2$ are bounded then we
have
\begin{equation*}
  \text{$x_1-x_2=c$\quad or\quad $c-1< x_1 -x_2 < c$\quad in $R$.}
\end{equation*}
If $y$ is unbounded then we have $y>\a_y$ in $R$.

For two distance graphs $G_1$, $G_2$ which are not necessarily in canonical
form, we denote by $\min(G_1,G_2)$ the distance graph where each edge
has the weight equal to the minimum of the corresponding weights in
$G_1$ and $G_2$.  Even though this graph may be not in canonical
form, it should be clear that it represents intersection of the two
arguments, that is, $\sem{\min(G_1,G_2)}=\sem{G_1}\cap\sem{G_2}$; in
other words, the valuations satisfying the constraints given by $\min(G_1,G_2)$
are exactly those satisfying all the constraints from $G_1$ as well as
 $G_2$.

We are now ready to examine the conditions when $R\cap Z$ is empty. We
start with the following simple lemma.
 
\begin{lemma}\label{lemma:bounded vars}
  Let $G_R$ be the distance graph of a region and let $x_1,x_2$ be two
  variables bounded in $R$. For every distance graph $G$: if in
  $\min(G_R,G)$ the weight of the edge $x_1\act{} x_2$ comes from $G$
  then $x_1\act{} x_2\act{} x_1$ is a negative cycle in $\min(G_R,G)$.
\end{lemma}

\begin{proof}
  Suppose that the edge $x_1\act{\fleq c} x_2$ is as required by the
  assumption of the lemma. In $R$ we can have either $x_2-x_1=d$ or
  $d-1< x_2-x_1<d$.

  In the first case we have edges $x_1\act{\leq d} x_2$ and
  $x_2\act{\leq -d} x_1$ in $G_R$. Since the edge $x_1\act{\fleq c}
  x_2$ comes from $G$ we have $c< d$ or $c=d$ and $\fleq$ is the
  strict inequality. We get a negative cycle $x_1\act{\fleq c} x_2
  \act{\leq -d} x_1$.

  In the second case we have edges $x_2\act{<-d+1} x_1$ and
  $x_1\act{<d} x_2$ in $R$. Hence $c<d$ and $x_1\act{\fleq c} x_2
  \act{<-d+1} x_1$ gives a negative cycle. \qed
\end{proof}

\subsubsection*{Proof of Proposition \ref{prop:intersection region
    zone}}

Let $G_R$, $G_Z$ be the canonical distance graphs representing the region $R$
and the zone $Z$ respectively.
One direction is immediate: If $\min(G_R,G_Z)$ has a negative
cycle then $R\cap Z$ is empty by Proposition~\ref{prop:cycles}.

For the other direction suppose that $R\cap Z$ is empty. Again, by
Proposition~\ref{prop:cycles} the graph $\min(G_R,G_Z)$ has a negative
cycle. An immediate case is when in this graph an edge between two
variables bound in $R$ comes from $G_Z$. From Lemma~\ref{lemma:bounded
  vars} we obtain a negative cycle on these two variables. So in what
follows we suppose that in $\min(G_R,G_Z)$ all the edges between
variables bounded in $R$ come from $G_R$. Hence every negative cycle
should contain an unbounded variable.

Let $y$ be a variable unbounded in $R$ that is a part of the negative
cycle. Consider $y$ with its successor and its predecessor on the
cycle: $x\act{}y\act{} x'$.  We will show that we can assume that $x'$
is $x_0$. Observe that in $G_R$ every edge to $y$ has value
$\infty$. So the weight of the edge $x\act{} y$ is from $G_Z$. If the
weight of the outgoing edge is also from $G_Z$ then we could have
eliminated $y$ from the cycle by choosing $x\act{} x'$ from
$G_Z$. Hence the weight of $y\act{} x'$ comes from $G_R$. Since $y$ is
unbounded in $R$, the weight of this edge is $d-\a_y$, where $d$ is
the value on the edge $0\act{\fleq d} x'$ in $G_R$. This is because we
can rewrite inequation $x'-y<d-\a_y$ as $y-x'>\a_y-d$, and we know
that $\a_y$ is the smallest possible value for $y$, while $d$ is the
supremum on the possible values of $x'$.  But then instead of the edge
$y\act{} x'$ we can take $y\act{} x_0 \act{} x'$ in $\min(G_R,G_Z)$
which has smaller value since we have $y\act{<-\a_y}x_0$ in $G_R$.

If $x$ is $x_0$ then we get a cycle of a required form since it
contains only $x_0$ and $y$. Otherwise, let us more closely examine
the whole negative cycle:
\begin{equation*}
  x_0\act{} x_{i_1}\act{} \dots \act{} x_{i_k}\act{} x \act{} y \act{}x_0\ .
\end{equation*}
By the reasoning from the previous paragraph, all of
$x_{i_1},\dots,x_{i_k}$ can be assumed to be bounded in $R$. Otherwise
we could get a cycle visiting $x_0$ twice and we could remove a part
of it with one unbounded variable and still have a negative cycle.  By
our assumption, all the edges from and to these variables come from
$R$. This means that the path from $x_0$ to $x$ can be replaced by an
edge $x_0\act{} x$ from $R$. So finally, the negative cycle has the
required form $x_0\act{} x\act{} y \act{} x_0$ with the edges $x_0
\xra{} x$ and $y \xra{} x_0$ coming from $G_R$ and the edge $x \xra{}
y$ coming from $G_Z$. Since $G_R$ is canonical, we can reduce this
cycle to $x \xra{} y \xra{} x$ with $x \xra{} y$ coming from $G_Z$ and
 $y \xra{} x$ coming from $G_R$.\qed

 \subsection{Efficient inclusion testing}

 Given two zones $Z$ and $Z'$ and a bound function $\a$, we would like
 to know if $Z \not\incl \Closure_\a(Z')$: that is, does there exist a
 region $R$ that intersects $Z$ but does not intersect $Z'$? From
 Proposition \ref{prop:intersection region zone} this reduces to
 asking if there exists a region $R$ that intersects $Z$ and two
 variables $x,y$ such that $Z'_{yx} + R_{xy} < (\le, 0)$. This brings
 us to look for the least value of $R_{xy}$ from among the regions $R$
 intersecting $Z$.  We begin with the observation that every
 consistent instantiation of an edge in a canonical distance graph $G$
 gives a valuation satisfying the constraints of $G$.

 \medskip

 \Repeat{Lemma}{lem:values} Let $G$ be a distance graph in canonical
 form, with all cycles positive. Let $x, y$ be two variables such that
 $x \act{\fleq_{xy} c_{xy}} y$ and $y \act{\fleq_{yx} c_{yx}} x$ are
 edges in $G$. Let $d \in \mathbb{R}$ such that $d \fleq_{xy} c_{xy} $
 and $-d \fleq_{yx} c_{yx} $. Then, there exists a valuation $v \in
 \sem{G}$ such that $v(y) - v(x) = d$.  \medskip

\begin{proof}
  Take $d$ as in the assumption of the lemma. Let $G_d$ be the
  distance graph where we have the edges $x \act{\le d} y$, and $y
  \act{ \le -d} x$ for variables $x$ and $y$ and the rest of the edges
  come from $G$. We show that all cycles in $G_d$ are positive. For
  contradiction, suppose there is a negative cycle $N$ in
  $G_d$. Clearly, since $G$ does not have negative cycle, $N$ should
  contain the variables $x$ and $y$. The value of the shortest path
  from $x$ to $y$ in $G$ was given by $(\fleq_{xy},
  c_{xy})$. Therefore, the shortest path value from $x$ to $y$ in
  $G_d$ is given by $d$ and the shortest path value from $y$ to $x$ is
  $-d$. Hence the sum of the weights in $N$ is negative would imply
  that the value of the cycle $x \rightarrow y \rightarrow x$ is
  negative. However since, this is $0$, such a negative cycle $N$
  cannot exist. The lemma follows from Proposition \ref{prop:cycles}.
  \qed
\end{proof}

Recall that for a zone $Z$, we denote by $Z_{xy}$ the weight of the
edge $x \xra{\fleq_{xy} c_{xy}} y$ in the canonical distance graph
representing $Z$.  We denote by $[v]$ the region to which $v$ belongs
to; $[v]_{xy}$ denotes the value $(\fleq_{xy}, r_{xy})$ of the
constraint $y - x \fleq_{xy} r_{xy}$ defining the region $[v]$. This
is precisely the value of the edge $x \xra{\fleq_{xy} r_{xy}} y$ in
the canonical distance graph representing the region $[v]$.  We are
interested in finding the least value of $[v]_{xy}$ from among the
valuations $v \in Z$. Lemmas \ref{lem:minimum-edge-region-0x-x0} and
\ref{lem:minimum-edge-region} describe this least value of $[v]_{xy}$
for different combinations of $x$ and $y$.

For a weight $(\fleq, c)$ we define $-(\fleq, c)$ as $(\fleq, -c)$. We
now define a \emph{ceiling} function $\ceil{\cdot}$ for weights.

\begin{definition}
  For a real $c$, let $\ceil{c}$ denote the smallest integer that is
  greater than or equal to $c$. We define the \emph{ceiling} function
  $\ceil{(\fleq, c)}$ for a weight $(\fleq, c)$ depending on whether
  $\fleq$ equals $\le$ or $<$, as follows:
 
\begin{equation*}
  \ceil{(\le,c)} = \begin{cases} (\le, c) & \text{if } c \text{ is an
      integer} \\
    (<, \ceil{c}) & \text{otherwise } \end{cases}
\end{equation*}

\begin{equation*}
  \ceil{(<,c)} = \begin{cases} (<, c+1) & \text{if } c \text{ is an
      integer} \\
    (<, \ceil{c}) & \text{otherwise } \end{cases}
\end{equation*}

\end{definition}

\begin{lemma}\label{lem:minimum-edge-region-0x-x0}
  Let $Z$ be a non-empty zone and let $x$ be a variable different from
  $x_0$. Then, from among the regions $R$ that intersect $Z$:

  \begin{itemize}
  \item the least value of $R_{0x}$ is given by
    \begin{equation*}
      \begin{cases}
        (<,\infty) & \text{if } Z_{x0} < (\le, -\a_x) \\
        \ceil{-Z_{x0}} & \text{otherwise}
      \end{cases}
    \end{equation*}

  \item the least value of $R_{x0}$ is given by $\max\{\ceil{-Z_{0x}},
    (<, -\a_x) \}$.

  \end{itemize}

\end{lemma}
\begin{proof}
  Let $Z_{0x} = (\fleq_{0x}, c_{0x})$ and $Z_{x0} = (\fleq_{x0},
  c_{x0})$.

  For the least value of $R_{0x}$, first note that if $Z_{x0} < (\le,
  -\a_x)$ then all valuations $v \in Z$ have $v_x > \a_x$ and by
  definition $[v]_{0x} = (<,\infty)$ for such valuations. If not, we
  know that for all valuations $-v_x \fleq_{x0} c_{x0}$, that is, $v_x
  \fgeq_{x0} -c_{x0}$. If $\fleq_{x0}$ is $\le$, then from Lemma
  \ref{lem:values} there exists a valuation with $v_x = -c_{x0}$ and
  this is the minimum value that can be attained. When $\fleq_{x0}$ is
  $<$, then we can find a positive $\epsilon < 1$ such that $c_{x0} -
  \epsilon \fleq_{x0} c_{x0}$ and $-c_{x0} + \epsilon \fleq_{0x}
  c_{0x}$. From Lemma \ref{lem:values}, there exists a valuation with
  $v_x = -c_{x0} + \epsilon$ for which $[v]_{0x} = (<, -c_{x0} +
  1)$. Since $\fleq_{x0}$ is a strict $<$, this is the minimum value
  for $R_{0x}$. This gives that the minimum value is $\ceil{-Z_{x0}}$.

  Now we look at the minimum value for $R_{x0}$. If
  $(\fleq_{0x},c_{0x}) \le (\le, \a_x)$, then all valuations $v$ have
  $v_x \le \a_x$ and by an argument similar to above, the minimum
  value of $R_{x0}$ would be given by $\ceil{-Z_{0x}}$. Since
  $(\fleq_{0x}, c_{0x}) \le (\le, \a_x)$, we have $(\fleq_{0x},
  -c_{0x}) \ge (\le, -\a_x) > (<, -\a_x)$.  If $(\fleq_{0x}, c_{0x}) >
  (\le, \a_x)$, then from Lemma \ref{lem:values}, there are valuations
  in $Z$ with $v_x > \a_x$ and for these valuations, $[v]_{x0} =
  (<,-\a_x)$. In this case the minimum value is given by
  $(<,-\a_x)$. Since $(\fleq_{0x}, c_{0x}) > (\le, \a_x)$, we have
  $(\fleq_{0x}, -c_{0x}) < (<, -\a_x)$ and so $\ceil{-Z_{0x}} \le
  (<,-\a_x)$.  In each case, observe that we get $\max\{
  \ceil{-Z_{0x}}, (<,-\a_x)\}$ as the minimum value.  \qed
\end{proof}
\medskip

\begin{lemma}\label{lem:minimum-edge-region}
  Let $Z$ be a non-empty zone and let $x,y$ be variables none of them
  equal to $x_0$. Then, from among the regions $R$ that intersect $Z$,
  the least value of $R_{xy}$ is given by
  \begin{equation*}
    \begin{cases}

      (<, \infty) & \text{ if } Z_{y0} < (\le, -\a_y) \\
      \max\{\ceil{-Z_{yx}}, \ceil{-Z_{y0}} - (\le, \a_x) \} & \text{
        otherwise}
    \end{cases}
  \end{equation*}
\end{lemma}

\begin{proof}
  Let $G$ be the canonical distance graph representing the zone
  $Z$. We denote the weight of an edge $i \xra{} j$ in $G$ by
  $(\fleq_{ij}, c_{ij})$. Recall that this means $Z_{ij} =
  (\fleq_{ij}, c_{ij})$. For clarity, for a valuation $v$, we write
  $v_x$ for $v(x)$.

  We are interested in computing the smallest value of the $y-x$
  constraint defining a region belonging to $\Closure_\a(Z)$, that is,
  we need to find $\min\{[v]_{xy} ~|~ v \in Z \}$. Call this $\b$. By
  definition of regions, we have for a valuation $v$:
  
  \begin{equation}\label{xy-value}
    [v]_{xy} = \begin{cases}
      (<,\infty) & \text{if } v_y > \a_y \\ 
      \ceil{(\le, v_y - v_x)} & \text{if } v_y \le \a_y \text{ and } v_x \le \a_x \\
      (<, \ceil{v_y} - \a_x) & \text{if } v_y \le \a_y \text{
        and } v_x > \a_x
    \end{cases}
  \end{equation}

  We now consider the first of the two cases from the statement of the
  lemma. Namely, $Z_{y0} < (\le, -\a_y)$. This means that
  $0-v_y\fleq_{y0} c_{y0}$ and $c_{y0}\leq -\a_{y}$; moreover
  $\fleq_{y0}$ is the strict inequality if $c_{y0}=-\a_y$. In
  consequence, all valuations $v\in Z$, satisfy $v_y > \a_y$. Whence
  $\b = (<, \infty)$.

  We now consider the case when $Z_{y0} \ge (\le, -\a_y)$.  Let $G'$
  be the graph in which the edge $0 \xra{} y$ has weight $\min\{ (\le,
  \a_y), (\fleq_{0y}, c_{0y})\}$ and the rest of the edges are the
  same as that of $G$. This graph $G'$ represents the valuations of
  $Z$ that have $v_y \le \a_y$: $\sem{G'} = \{ v \in Z ~|~ v_y\le
  \a_y\}$. We show that this set is not empty. For this we check that
  $G'$ does not have negative cycles. Since $G$ does not have negative
  cycles, every negative cycle in $G'$ should include the newly
  modified edge $0 \xra{} y$. Note that the shortest path value from
  $y$ to $0$ does not change due to this modified edge. So the only
  possible negative cycle in $G'$ is $0 \xra{} y \xra{} 0$. But then
  we are considering the case when $Z_{y0} \ge (\le, -\a_y)$, and so
  $Z_{y0} + (\le, \a_y) \ge (\le,0)$. Hence this cycle cannot be
  negative either. In consequence all the cycles in $G'$ are positive
  and $\sem{G'}$ is not empty.

  To find $\b$, it is sufficient to consider only the valuations in
  $\sem{G'}$. As seen from Equation \ref{xy-value}, among the
  valuations in $\sem{G'}$, we need to differentiate between those
  with $v_x \le \a_x$ and the ones with $v_x > \a_x$.  We proceed as
  follows. We first compute $\min\{ [v]_{xy}~|~v \in \sem{G'} \text{
    and } v_x \le \a_x\}$. Call this $\b_1$. Next, we compute
  $\min\{[v]_{xy}~|~v \in \sem{G'} \text{ and } v_x > \a_x\}$ and set
  this as $\b_2$. Our required value $\b$ would then equal
  $\min\{\b_1, \b_2\}$.

  To compute $\b_1$, consider the following distance graph $G'_1$
  which is obtained from $G'$ by just changing the edge $0 \xra{} x$
  to $\min\{ (\le, \a_x), (\fleq_{0x}, c_{0x}) \}$ and keeping the
  remaining edges the same as in $G'$. The set of valuations
  $\sem{G'_1}$ equals $\{ v \in \sem{G'} ~|~ v_x \le \a_x \}$. % Since
  % $G$ does not have negative cycles, a negative cycle in the graph
  % $G'_1$ should be either due to the edge $0 \xra{} y$ or the edge
  % $0
  % \xra{} x$. Both of them cannot occur in the same negative
  % cycle. Also, note that the shortest path from either $x$ or $y$ to
  % $0$ does not change due to these two newly modified
  % edges. Therefore, the possible negative cycles are $0 \xra{} y
  % \xra{} 0$ and $0 \xra{} x \xra{} 0$. In the former case,
  % $\sem{G'}$
  % would be empty from Equation \ref{G'-empty}. Hence we get:
  % \begin{equation}
  %   \begin{array}{ccc}
  %     \sem{G'_1} = \es &\text{iff}& \text{either }  G' = \es \text{ or} \\
  %     & & (\le, \a_x) < Z_{0x} \text{ and } (\le, \a_x) + Z_{x0} < (\le, 0)\\
  %   \end{array}
  % \end{equation}
  If $\sem{G'_1} = \es$, we set $\b_1$ to $(<, \infty)$ and proceed to
  calculate $\b_2$. If not, we see that from Equation \ref{xy-value},
  for every $v \in \sem{G'_1}$, $[v]_{xy}$ is given by $\ceil{(\le,
    v_y - v_x)}$. Let $(\fleq_1,w_1)$ be the shortest path from $y$ to
  $x$ in the graph $G'_1$. Then, we have for all $v \in \sem{G'_1}$,
  $v_x - v_y \fleq_1 w_1$, that is, $ v_y - v_x \fgeq_1 -w_1$. If
  $\fleq_1$ is $\le$, then the least value of $[v]_{xy}$ would be
  $(\le, -w_1)$ and if $\fleq_1$ is $<$, one can see that the least
  value of $[v]_{xy}$ is $(<, -w_1+1)$. This shows that $\b_1 =
  \ceil{(\fleq_1, -w_1)}$. It now remains to calculate $(\fleq_1,
  w_1)$.

  Recall that $G'_1$ has the same edges as in $G$ except possibly
  different edges $0 \xra{} x$ and $0 \xra{} y$. If the shortest path
  from $y$ to $x$ has changed in $G'_1$, then clearly it should be due
  to one of the above two edges. However note that the edge $0 \xra{}
  y$ cannot belong to the shortest path from $y$ to $x$ since it would
  contain a cycle $y \xra{} \dots 0 \xra{} y \xra{} \dots x$ that can
  be removed to give shorter path. Therefore, only the edge $0 \xra{}
  x$ can potentially yield a shorter path: $y \xra{} \dots \xra{} 0
  \xra{} x$. However, the shortest path from $y$ to $0$ in $G'_1$
  cannot change due to the added edges since that would form a cycle
  with $0$ and we know that all cycles in $G'_1$ are positive.
  Therefore the shortest path from $y$ to $0$ is the direct edge $y
  \xra{} 0$, and the shortest path from $y$ to $x$ is the minimum of
  the direct edge $y \xra{} x$ and the path $y \xra{} 0 \xra{} x$. We
  get: $(\fleq_1, w_1) = \min\{ (\fleq_{yx}, c_{yx}), (\fleq_{y0},
  c_{y0}) + (\le, \a_x) \}$ which equals $\min\{ Z_{yx}, Z_{y0} +
  (\le, \a_x)\}$. Finally, from the argument in the above two
  paragraphs, we get:

\begin{equation}\label{b1}
  \b_1 = \begin{cases}
    (<, \infty) & \text{ if } \sem{G'_1} = \es \\
    \ceil{-Z_{yx}} & \text{ if } \sem{G'_1} \neq \es \text{ and } Z_{yx}
    \le Z_{y0} + (\le, \a_x) \\
    \ceil{-Z_{y0}} + (\le, -\a_x) & \text{ if } \sem{G'_1} \neq \es
    \text{ and } Z_{yx} > Z_{y0} + (\le, \a_x) \\
  \end{cases}
\end{equation}

We now proceed to compute $\b_2=\min\{ [v]_{xy} ~|~ v \in \sem{G'}
\text{ and } v_x > \a_x \}$. Let $G'_2$ be the graph which is obtained
from $G'$ by modifying the edge $x \xra{} 0$ to $\min\{ Z_{x0},
(<,-\a_x)\}$ and keeping the rest of the edges the same as in
$G'$. Clearly $\sem{G'_2} = \min\{ v \in \sem{G'}~|~ v_x >
\a_x\}$. % Again,
% since $G$ does not have negative cycles, if $G'_2$ has a negative
% cycle, it would be due to one of the modified edges, either $0
% \xra{} y$ or $x \xra{} 0$. This gives the following three potential
% negative cycles in $G'_2$: $0 \xra{} y \xra{} 0$, $0 \xra{} y \xra{}
% x \xra{} 0$, $0 \xra{} x \xra{} 0$. If $0 \xra{} y \xra{} 0$ is
% negative, then it implies that the weight on the edge $0 \xra{} y$
% is $(\le, \a_y)$. From Equation \ref{G'-empty}, we get that
% $\sem{G'}$ is empty in this case. We now look at the two other
% potential negative cycles. Let us denote by $(\fleq, w)$ the
% shortest path from $0$ to $x$ in $G'_2$. We get $(\fleq, w) = \min\{
% Z_{0x}, (\le, \a_y) + Z_{yx}\}$. Saying either $0 \xra{} y \xra{} x
% \xra{} 0$ or $0 \xra{} x \xra{} 0$ is negative reduces to saying
% $(\fleq, w) + (<, -\a_x) < (\le, 0)$. Therefore, we get:
 
% \begin{equation}
%   \begin{array}{ccc}
%     \sem{G'_2} = \es & \text{iff} & \text{either } G' = \es \text{ or} \\
%     & & (<, -\a_x) < Z_{x0} \text{ and } (<,-\a_x) + (\fleq, w) < (\le, 0) \\
%   \end{array}
% \end{equation}

Again, if $\sem{G'_2}$ is empty, we set $\b_2$ to
$(<,\infty)$. Otherwise, from Equation \ref{xy-value}, for each
valuation $v \in \sem{G'_2}$, the value of $[v]_{xy}$ is given by
$(<,\ceil{v_y} - \a_x)$. For the minimum value, we need the least
value of $v_y$ from $v \in \sem{G'_2}$. Let $(\fleq_2, w_2)$ be the
shortest path from $y$ to $0$ in $G'_2$. Then, since $-v_y \fleq_2
w_2$, the least value of $\ceil{v_y}$ would be $-w_2$ if $\fleq_2 =
\le$ and equal to $\ceil{-w_2}$ if $\fleq_2 = <$ and $\b_2$ would
respectively be $(<, -w_2 -\a_x)$ or $(<, -w_2 + 1 -\a_x)$. It now
remains to calculate $(\fleq_2, w_2)$.

Recall that $G'_2$ is $G$ with $0 \xra{} y$ and $x \xra{} 0$
modified. The shortest path from $y$ to $0$ cannot include the edge $0
\xra{} y$ since it would need to contain a cycle, for the same reasons
as in the $\b_1$ case.  So we get $(\fleq_2, w_2) = \min\{ Z_{y0},
Z_{yx} + (<,-\a_x)\}$. If $Z_{y0} \le Z_{yx} + (<,-\a_x)$, then we
take $(\fleq_2, w_2)$ as $Z_{y0}$, otherwise we take it to be $Z_{yx}
+ (<, -\a_x)$. So, we get $\b_2$ as the following:
% We want to write $\b_2$ in the same form as $\b_1$ (Equation
% \ref{b1}) so that we can write $\min\{\b_1, \b_2\}$ conveniently. To
% this regard, note the following equalities:

% \begin{align*}
%   & Z_{y0} < Z_{yx} + (<,-\a_x) \\
%   \iff~ & Z_{y0} + (<,\a_x) < Z_{yx} \\
%   \iff~ & Z_{y0} + (\le, \a_x) \le Z_{yx}
% \end{align*}

% Summarizing, we get:

\begin{equation}\label{b2}
  \b_2 = \begin{cases}
    (<, \infty) & \text{ if } \sem{G'_2} = \es \\
    -Z_{yx} + (<,1) & \text{ if } \sem{G'_2} \neq \es \text{ and }
    Z_{y0} \ge Z_{yx} + (<,-\a_x) \\
    \ceil{-Z_{y0}} + (<, -\a_x) & \text{ if } \sem{G'_2} \neq \es
    \text{ and } Z_{y0} < Z_{yx} + (<,
    -\a_x) \\
  \end{cases}
\end{equation}
However, we would like to write $\b_2$ in terms of the cases used for
$\b_1$ in Equation \ref{b1} so that we can write $\b$, which equals
$\min\{ \b_1, \b_2\}$, conveniently.

Let $\p_1$ be the inequation: $Z_{yx} \le Z_{y0} + (\le, \a_x)$. From
Equation \ref{b1}, note that $\b_1$ has been classified according to
$\p_1$ and $\neg \p_1$ when $\sem{G'_1}$ is not empty. Similarly, let
$\p_2$ be the inequation: $Z_{y0} \ge Z_{yx} + (<,-\a_x)$. From
Equation \ref{b2} we see that $\b_2$ has been classified in terms of
$\p_2$ and $\neg \p_2$ when $\sem{G'_2}$ is not empty. Notice the
subtle difference between $\p_1$ and $\p_2$ in the weight component
involving $\a_x$: in the former the inequality associated with $\a_x$
is $\le$ and in the latter it is $<$. This necessitates a bit more of
analysis before we can write $\b_2$ in terms of $\p_1$ and $\neg
\p_1$.

Suppose $\p_1$ is true. So we have $(\fleq_{yx},c_{yx}) \le
(\fleq_{y0}, c_{y0} + \a_x)$. This implies: $c_{yx} \le c_{y0} +
\a_x$. Therefore, $c_{y0} \ge c_{yx} - \a_x$. When $c_{y0} > c_{yx} -
\a_x$, $\p_2$ is clearly true. For the case when $c_{y0}= c_{yx} -
\a_x$, note that in $\p_2$ the right hand side is always of the form
$(<, c_{yx} - \a_x)$, irrespective of the inequality in $Z_{yx}$ and so
yet again, $\p_2$ is true. We have thus shown that $\p_1$ implies
$\p_2$.  %  Hence when $\p_1$ is true, that is, when
% $Z_{yx} \le Z_{y0} + (\le,\a_x)$, we can take $\b_2$ to be $-Z_{yx} +
% (<,1)$.

Suppose $\neg \p_1$ is true.  We have $(\fleq_{yx}, c_{yx}) >
(\fleq_{y0}, c_{y0} + \a_x)$. If $c_{yx} > c_{y0} +
\a_x$, then clearly $c_{y0} < c_{yx} - \a_x$ implying that $\neg \p_2$
holds. If $c_{yx} = c_{y0} + \a_x$, then we need to
have $\fleq_{yx} = \le$ and $\fleq_{y0} = <$. Although $\neg \p_2$
does not hold now, we can safely take $\b_2$ to be
$\ceil{-Z_{y0}} + (<, -\a_x)$ as its value is in fact equal to
$-Z_{yx} + (<,1)$ in this case. Summarizing the above two paragraphs,
we can rewrite $\b_2$ as follows:

\begin{equation}\label{b2-final}
  \b_2 = \begin{cases}
    (<, \infty) & \text{ if } \sem{G'_2} = \es \\
    -Z_{yx} + (<,1) & \text{ if } \sem{G'_2} \neq \es \text{ and }
    Z_{xy} \le Z_{y0} + (\le,\a_x) \\
    \ceil{-Z_{y0}} + (<, -\a_x) & \text{ if } \sem{G'_2} \neq \es
    \text{ and } Z_{xy} > Z_{y0} + (\le, \a_x) \\
  \end{cases}
\end{equation}

We are now in a position to determine $\b$ as $\min\{\b_1, \b_2
\}$. Recall that we are in the case where $Z_{y0} \le (\le, -\a_y)$ and we have
established that $\sem{G'}$ is non-empty. Now since $\sem{G'} =
\sem{G'_1} \cup \sem{G'_2}$ by construction, both of them cannot be
simultaneously empty. Hence from Equations \ref{b1} and
\ref{b2-final}, we get $\b$, the $\min\{\b_1, \b_2 \}$ as:

\begin{equation}\label{b}
  \b_ = \begin{cases}
    \ceil{-Z_{yx}} & \text{ if } Z_{xy} \le Z_{y0} + (\le,\a_x) \\
    \ceil{-Z_{y0}} + (<, -\a_x) & \text{ if } Z_{xy} > Z_{y0} + (\le, \a_x) \\
  \end{cases}
\end{equation}

% If the constant $c_{yx} < c_{y0} + \a_x$, then clearly the
% inequation $Z_{y0} > Z_{yx} + (<, -\a_x)$ holds; in other words,
% $\neg \p_2$ is true and hence $\b_2$ is $-Z_{yx} + (<,1)$ from
% Equation \ref{b2}. If the constant $c_{yx} = c_{y0} + \a_x$ and if
% $\fleq_{y0} = \le$, then again $\neg \p_2$ is true; if $\fleq_{y0} =
% <$, then $\p_1$ implies that $\fleq_{yx} = <$ too.

% We first prove that $\psi_2 \imp \neg \psi_1$. The inequation
% $\psi_2$ says that $(\fleq_{y0}, c_{y0}) \le (<, c_{yx} -
% \a_x)$. When the constant $c_{y0} < c_{yx} - \a_x$, then clearly
% $\neg \psi_1$ holds. When $c_{y0} = c_{yx} - \a_x$, then $\p_2$
% being true implies that $\fleq_{y0} = <$. If $\fleq_{yx} = <$ too,
% clearly $\neg \p_1$ is true. However, when $\fleq_{yx} = \le$, $\neg
% \p_1$ no longer holds.

There remains one last reasoning. To prove the lemma, we need to show
that $\b = \max\{ \ceil{-Z_{yx}}, \ceil{-Z_{y0}} + (<,-\a_x)\}$.  For
this it is enough to show the following two implications:

\begin{align*}
  Z_{yx} \le Z_{y0} + (\le, \a_x) \imp \ceil{-Z_{yx}} \ge
  \ceil{-Z_{y0}} + (<, -\a_x) \\
  Z_{yx} > Z_{y0} + (\le, \a_x) \imp \ceil{-Z_{yx}} \le
  \ceil{-Z_{y0}}+(<,-\a_x)
\end{align*}
We prove only the first implication. The second follows in a similar
fashion.  Let us consider the notation
$(\fleq_{yx}, c_{yx})$ and $(\fleq_{y0}, c_{y0})$ for $Z_{yx}$ and
$Z_{y0}$ respectively.
 So we have:

\begin{align*}
  & (\fleq_{yx}, c_{yx}) \le (\fleq_{y0},c_{y0}) + (\le, \a_x) \\
  \imp~ & (\fleq_{yx}, c_{yx}) \le (\fleq_{y0}, c_{y0} + \a_x)
\end{align*}
If the constant $c_{yx} < c_{y0} + \a_x$, then $-c_{yx} > -c_{y0} -
\a_x$ and we clearly get that $\ceil{-Z_{yx}} \ge \ceil{-Z_{y0}} +
(<,-\a_x)$. If the constant $c_{yx} = c_{y0} + \a_x$ and if
$\fleq_{y0} = \le$, then the required inequation is trivially true; if
$\fleq_{y0} = <$, it implies that $\fleq_{yx} = <$ too and clearly
$\ceil{(<,-c_{yx})}$ equals $ \ceil{(<,-c_{y0})} + (<,-\a_x)$. \qed

% then $\fleq_{y0}$ should be $\le$ and $\fleq_{yx}$ should be $<$ and
% we clearly see that $\ceil{(<,-c_{yx})}$, that is $\ceil{-Z_{yx}}$
% is greater than $\ceil{-Z_{y0}} + (<,-\a_x)$.  \qed
\end{proof}

\medskip

\Repeat{Theorem}{thm:2clocks}Let $Z, Z'$ be zones. Then, $Z \nsubseteq
\Closure_\a(Z')$ iff there exist variables $x$, $y$ such that one of
the following conditions hold:
\begin{enumerate}

\item $Z'_{0x} < Z_{0x} \text{ and } Z'_{0x} \le (\a_x, \le)$, or
\item $Z'_{x0} < Z_{x0} \text{ and } Z_{x0} \ge (-\a_x, \le)$, or
\item $Z_{x0} \ge (-\a_x, \le) \text{ and } Z'_{xy} < Z_{xy} \text{
    and } Z'_{xy} \le (\a_y, \le) + \floor{Z_{x0}} $
\end{enumerate}
\medskip

\begin{proof}
  By definition of the $\Closure$ abstraction, $Z \nsubseteq
  \Closure_\a(Z')$ iff there exists a region $R$ that intersects $Z$
  but does not intersect $Z'$. Therefore, from Proposition
  \ref{prop:intersection region zone}, we need an $R$ that intersects
  $Z$ and satisfies $Z'_{yx} + R_{xy} < (\le, 0)$ for some variables
  $x,y$. This is equivalent to saying that for the least value of
  $R_{xy}$ that can be obtained from the zone $Z$, we have $Z'_{yx} +
  R_{xy} < (0,\le)$. Depending on if $x$ is $x_0$ or $y$ is $x_0$ or
  both $x$ and $y$ are not $x_0$ we get the following three conditions
  that correspond to the three conditions given in the theorem.

  \begin{description}
  \item{Case 1:} $Z'_{0x} + R_{x0} < (\le,0)$ \\
    From Lemma \ref{lem:minimum-edge-region-0x-x0}, the minimum value
    of $R_{x0}$ from among the regions intersecting $Z$ is given by
    $\max\{ \ceil{-Z_{0x}},(<, -\a_x)\}$. So we have:
    \begin{align*}
      Z'_{0x} & + \max\{ \ceil{ -Z_{0x}}, (<, -\a_x)\} < (\le, 0)  \\
      & \imp ~ Z'_{0x} + \ceil{ -Z_{0x} } < (\le,0)~ \text{ and } ~
      Z'_{0x} + (<,
      -\a_x) < (\le,0) \\
      & \imp ~ Z'_{0x} < Z_{0x} ~ \text{ and } ~ Z'_{0x} \le (\le,
      \a_x)
    \end{align*}
    This gives Condition 1 of the theorem.
  \item{Case 2:} $Z'_{x0} + R_{0x} < (\le,0)$ \\
    From Lemma \ref{lem:minimum-edge-region-0x-x0}, the minimum value
    of $R_{0x}$ is $(<,\infty)$ if $-Z_{x0} > (\le, \a_x)$ and hence
    it cannot be part of a negative cycle. The edge $R_{0x}$ can yield
    a negative cycle only when $-Z_{x0} \le (\le, \a_x)$, in which
    case the least value of $R_{0x}$ is given by $\ceil{-Z_{x0}}$. So
    we have $Z'_{x0} + \ceil{-Z_{x0}} < (\le,0)$ which translates to
    $Z'_{x0} < Z_{x0}$. Therefore, this case is equivalent to saying
    $Z_{x0} \ge (\le, -\a_x)$ and $Z'_{x0} < Z_{x0}$ which gives
    Condition 2 of the theorem.

  \item{Case 3:} $Z'_{yx} + R_{xy} < (\le,0)$ From Lemma
    \ref{lem:minimum-edge-region}, we get that the minimum value of
    the edge $R_{xy}$ is $(<, \infty)$ if $-Z_{y0} > (\le,
    \a_y)$. Similar to the case above, $R_{xy}$ cannot be part of a
    negative cycle if $-Z_{y0} > (\le, \a_y)$. So we need to first
    check if $-Z_{y0} \le (\le, \a_y)$, that is, if $Z_{y0} \ge (\le,
    -\a_y)$. Now, from Lemma \ref{lem:minimum-edge-region}, the
    minimum value of $R_{xy}$ is given by the $\max\{ \ceil{-Z_{yx}},
    \ceil{-Z_{y0}} - (\le, \a_x) \}$. We get:

    \begin{align*}
      Z'_{yx} & + \max\{\ceil{-Z_{yx}},
      \ceil{-Z_{y0}} - (\le, \a_x)  \} < (\le, 0) \\
      & \imp Z'_{yx} + \ceil{-Z_{yx}} < (\le, 0) \text{ and } Z'_{yx}
      +
      \ceil{-Z_{y0}} -(\le, \a_x) < (\le, 0) \\
      & \imp Z'_{yx} < Z_{yx} \text{ and } Z'_{yx} + \ceil{-Z_{y0}} -
      (\le, -\a_x) < (\le,0)
    \end{align*}

    Let us look at the second inequality: $Z'_{yx} + \ceil{-Z_{y0}} -
    (\le, -\a_x) < (\le,0)$. If $Z_{y0}$ is of the form $(\le,c)$ with
    $c$ an integer, then $-Z_{y0} = (\le, -c)$ and $\ceil{-Z_{y0}}$ is
    the same: $(\le, -c)$. So we get:

    \begin{align*}
      Z'_{yx} + (\le, -c) + (<, -\a_x) < (\le,
      0) \\
      \iff Z'_{yx} + (<, -c - \a_x) < (\le, 0) \\
      \iff Z'_{yx} \le (\le, c+\a_x) \\
    \end{align*}
    
    When $Z_{y0} = (<,c)$, then $\ceil{-Z_{y0}} = (<, -c+1)$ and we
    get:
    
    \begin{align*}
      Z'_{yx} + (<, -c+1) + (<, -\a_x) < (\le, 0) \\
      \iff Z'_{yx} + (<, -c+1-\a_x) < (\le, 0) \\
      \iff Z'_{yx} \le (\le, c-1 +\a_x) \\
    \end{align*}
    This gives Condition 3 of the Theorem (symmetric in $x$ and $y$).
  \end{description}
  \qed
\end{proof}

\subsection{Handling LU-approximation}

Recall that for a zone $Z$, we denote by $Z^+$ the zone
$\elup{Z}$. Also note that $Z^+$ is not necessarily in canonical
form. 

\begin{proposition}\label{prop:LU-negative-cycle}
  Let $R$ be a region and $Z$ be a zone. Then, $R \cup Z^+$ is empty
  iff there exist variables $x,y$ such that $Z^+_{yx} + R_{xy} < (\le,
  0)$. 
\end{proposition}

\begin{proof}
  Let $G_R$ be the canonical graph representing $R$ and let $G_Z$ be
  the canonical distance graph representing $Z$. Let $G_{Z^+}$ be the
  graph that representing $Z^+$. By definition, $G_{Z^+}$ is obtained
  from $G_Z$ by changing some edges to $(<,\infty)$ and some edges
  incident on $x_0$ to $(<,-U(x))$. Also, note that $G_{Z^+}$ is not
  necessarily in canonical form.
 
  From Proposition \ref{prop:cycles}, $R \cup Z^+$ is empty iff
  $\min(G_R, G_{Z^+})$ has a negative cycle. An easy case is when in
  $\min(G_R, G_{Z^+})$ a weight of an edge between two variables bound
  in $R$ comes from $G_{Z^+}$.  Using Lemma~\ref{lemma:bounded vars}
  we get a negative cycle of the required form on these two variables.

  It remains to consider the opposite case. We need then to have an
  unbounded variable on the cycle.  Let $y$ be a variable unbounded in
  $R$ that is part of the negative cycle. Consider $y$ with its
  successor and its predecessor on the cycle: $x\act{}y\act{} x'$.
  Observe that in $R$ every edge to $y$ has value $\infty$. So the
  weight of the edge $x\act{} y$ is from $Z^+$. By definition of
  $Z^+$, it is also from $Z$.  If also the weight of the outgoing edge
  were from $Z$ then we could have obtained a shorter negative cycle
  by choosing $x\act{} x'$ from $Z$. Hence the weight of $y\act{} x'$
  comes from an edge modified in $Z^+$ or from $R$. In the first case
  it is $y\act{<-U(y)}0$, in the second it is
  $y\act{<-\a_y}0$. However, note that since $U(y) \le \a_y$, we have
  $ -\a_y \le -U(y)$ and therefore, in $\min(G_R, G_{Z^+})$ we could
  consider the edge to come from $R$, that is $y \act{< -\a_y} 0$.

  The same analysis as in the proof of
  Proposition~\ref{prop:intersection region zone} we get that the
  shortest cycle of this kind should be of the form $ 0\act{} y \act{}
  0$ or $0\act{} x\act{} y\act{} 0$; where $y$ is an unbound variable
  and $x$ is a bound variable. This cycle has the required form.\qed
\end{proof}

% \begin{corollary}\label{cor:luconditions}
%   Let $R$ be a region and let $Z$ be a zone. Let $Z^+ =
%   \Extra_{LU}^+(Z)$.  The intersection $R \cap Z^+$ is empty iff one
%   of the following conditions holds (below $x,x'$ are variables bound
%   in $R$ and $y$ a variable unbound in $R$):

%   \begin{enumerate}
%   \item $Z^+_{0x}< R_{0x}$, or
%   \item $Z^+_{x0}< R_{x0}$, or
%   \item $Z^+_{xx'}< R_{xx'}$, or
%   \item $Z^+_{0y} \le (\a_y, \le)$, or
%   \item $Z^+_{xy} \le (\a_y -c, \le)$ where $c$ is the constant
%     appearing in $R_{0x}$.
%   \end{enumerate}
% \end{corollary}

\subsubsection{Efficient inclusion testing for LU approximations}

% The conditions for a region $R$ to not intersect the extrapolated zone
% $Z^+$ are obtained from Lemma \ref{lemma:LU-negative-cycle}. It is
% similar to the condition for non-intersection of $R$ with $Z$ as given
% in Proposition \ref{prop:intersection region zone} however with $Z$
% replaced by $Z^+$.

Let $Z, Z'$ be two zones and let $G_Z, G_{Z'}$ be the respective
distance graphs in canonical form. By extrapolating $G_{Z'}$ with
respect to the $\Extra_{LU}^+$ operator gives a zone $Z^+$ and a
corresponding distance graph $G_{Z^+}$, which is not necessarily in
canonical form. However, from Proposition~\ref{prop:LU-negative-cycle},
the check $Z \subseteq \Closure_{LU}^+(Z')$ can be reduced to an edge
by edge comparison with every region intersecting
$Z$. Lemmas \ref{lem:minimum-edge-region-0x-x0} and
\ref{lem:minimum-edge-region} give the least value of the edge
$R_{xy}$ for a region intersecting $Z$. Hence, similar
to the case of $Z \subseteq \Closure_\a(Z')$, it is enough to look at
edges of $G_Z$ one by one to look at what regions we can possibly
get. As a result we get an analogue of Theorem \ref{thm:2clocks}
with $Z'$ replaced by $Z'^+$.

\begin{theorem} 
  Let $Z, Z'$ be zones. Writing $Z'^+$ for $\Extra^+_{LU}(Z')$ we get
  that $Z \nsubseteq \Closure_\a(Z'^+)$ iff there exist variables $x$,
  $y$ such that one of the following conditions hold:

  \begin{enumerate}
  \item $Z'^+_{0x} < Z_{0x} \text{ and } Z'^+_{0x} \le (\a_x, \le)$, or
  \item $Z'^+_{x0} < Z_{x0} \text{ and } Z_{x0} \ge (-\a_x, \le)$, or
  \item $Z_{x0} \ge (-\a_x, \le) \text{ and } Z'^+_{xy} < Z_{xy}
    \text{ and }
    Z'^+_{xy} \le (\a_y, \le) + Z_{x0} $
  \end{enumerate}
\end{theorem}

\section{Proofs from Section 4}

\subsection{Correctness of the algorithm with $\Closure$ approximation}

Here we show the proof of 

\textbf{Theorem~\ref{thm:alg correct}}
  An accepting state is reachable in $ZG(\Aa)$ iff the
  algorithm reaches a node with an accepting state and a non-empty zone. 
\medskip

The right-to-left direction follows by a straightforward induction on
the length of the path. The left-to-right direction is shown using the
following lemmas.
 
Let $Post(S,t)$ stand for the set of all valuations of clocks
reachable by $t$ from valuations in $S$. We will need the following
observation.
% made in \cite{Bouyer:FMSD:2004} and provide a proof in the
%Appendix \ref{app:post-closure}.

\begin{lemma}[\cite{Bouyer:FMSD:2004}]\label{lem:closurepost}
  For every zone $Z$, transition $t$ and a bound function $\a$:
  \begin{equation*}
    Post(Closure_\a(Z),t) \incl Closure_\a(Post(Z,t)).    
  \end{equation*}
\end{lemma}

\begin{lemma}\label{zone-lazy}
  Suppose that algorithm concludes that the final state is not
  reachable. Consider the tree it has constructed.  For every $(q,Z)$
  reachable from $(q_0,Z_0)$ in $ZG(\Aa)$, there is a non tentative
  node $(q,Z',\a')$ in the tree, such
  that $Z \subseteq Closure_{\a'}(Z')$.
\end{lemma}

\begin{proof}
  The hypothesis is vacuously true for $(q_0,Z_0)$.  Assume that the
  hypothesis is true for a node $(q,Z) \in ZG(\Aa)$. We now prove that
  the lemma is true for every successor of $(q,Z)$.

  From hypothesis, there exists a non tentative node $(q,Z_L,\a)$ in
  the constructed tree such that $Z \subseteq Closure_\a(Z_L)$. Let
  $t= (q, g, r, q')$ be a transition of $\Aa$ and let $(q,Z)
  \xrightarrow{t}(q',Z') \in ZG(\Aa)$.
 
  The transition $t$ is enabled from $(q,Z_L,\a)$ because $Z \subseteq
  Closure_\a(Z_L)$, and, due to constraint propagation, for every
  clock $x$, $\a_x$ is greater than the maximum constant it is
  compared to in the guard $g$. So we have
  \begin{equation*}
    (q,Z_L,\a) \xrightarrow{t}
    (q',Z_L',\a')
  \end{equation*}
  in the constructed tree.

  Since $Z \incl \Closure_\a(Z_L)$, we have $Post(Z,t)\incl
  Post(Closure_{\a}(Z_L),t)$, that is $Z' \incl
  Post(Closure_{\a}(Z_L),t)$.  From Lemma \ref{lem:closurepost}, $Z'
  \incl Closure_\a(Post(Z_L,t))$, that is $Z' \incl
  Closure_\a(Z_L')$. We now need to check if we can replace $\a$ with
  $\a'$. But $Closure_\a(Z'_L)\incl Closure_{\a'}(Z_L')$ since by
  definition of constant propagation $\a_x\geq \a'(x)$ for all clocks
  $x$ not reset by $t$, and for clocks $x$ that are reset, $Z_L'$
  entails $x = 0$, therefore irrespective of $\a$ or $\a'$ the regions
  that intersect with $Z_L'$ should satisfy $x = 0$.

  If $n'=(q',Z_L',\a') $ is non tentative, we are done and $n'$ is the
  node in the constructed tree corresponding to $(q',Z')$. If $n'$ is
  tentative then by definition we know that there exists a non tentative
  node $(q',Z_L'',\a'')$ such that $\a''=\a'$ and $Z_L' \incl
  Closure_{\a'}(Z_L'')$. Thus $Z' \incl Closure_{\a'}(Z_L'')$. In this
  case $(q',Z_L'', \a'')$ is the node corresponding to $(q',Z')$.

  \qed
\end{proof}

\subsection{Correctness of the algorithm with LU approximation}\label{sec:LU-correct}

The proof of the correctness of the algorithm using $Z\incl
\Closure_{LU}^+(Z')$ test is similar to that using $Z\incl
\Closure_\a(Z')$ test. We call it LU-algorithm for short. Since
$\Extra_{LU}^+$ is difficult to handle, we do a small detour through
another approximation $\Alu(Z)$ introduced
in~\cite{Behrmann:STTT:2006}. We recall its definition here.

\begin{definition}[LU-preorder]
  Fix integers $L$ and $U$. Let $\nu$ and $\nu'$ be two
  valuations. Then, we say $\nu' \lu \nu$ if for each clock $x$:
  \begin{itemize}
  \item either $\nu'(x) = \nu(x)$,
  \item or $L(x) < \nu'(x) < \nu(x)$,
  \item or $U(x) < \nu(x) < \nu'(x)$.
  \end{itemize}
\end{definition}

This LU-preorder can be extended to define abstractions of sets of
valuations.

\begin{definition}[$\Alu$, abstraction w.r.t $\lu$]
  Let $W$ be a set of valuations. Then,
  \begin{equation*}
    \Alu(W) = \{ \nu ~|~ \exists \nu' \in W, ~\nu' \lu \nu \}
  \end{equation*}
\end{definition}

It is shown in \cite{Behrmann:STTT:2006} that this is a sound, complete
and finite abstraction, coarser than $\Closure$. The soundness of this
abstraction follows from the lemma given below.

\begin{lemma}\label{lem:a-simulates}
  Let $q$ be a state of $\Aa$ and $t = (q,g,R,q_1)$ a
  transition. Assume that for a clock $x$: $L(x) \ge c$ for all $c$
  such that $x \ge c$ or $x >c$ occurs in $g$; and $U(x) \ge d$ for
  all $d$ such that $x \le d$ or $x < d$ occurs in $g$.  Let $\nu$ and
  $\nu'$ be valuations such that $\nu' \lu \nu$.  Then, $(q,\nu)
  \xrightarrow{\delta, t} (q_1, \nu_1)$ implies that there exists a
  delay $\d'$ and a valuation $\nu'_1$ such that $(q,\nu')
  \xrightarrow{\delta', t} (q_1, \nu'_1)$ and $\nu'_1 \lu \nu_1$.
\end{lemma}

The relation between $\Alu(Z)$ and $\Extra^+_{LU}(Z)$ is summarized by
the following.

\begin{lemma}\label{lem:recall}
  For all zones $Z$,
  \begin{align}
    \elup{Z} \text{ is a zone}  \\
    \elup{Z} \subseteq \Alu(Z)
  \end{align}
\end{lemma}

We are now in a position to prove the correctness of LU-algorithm.

\begin{theorem}
  An accepting state  is reachable in $ZG(\Aa)$ iff the
  LU-algorithm reaches a node with an accepting state and a non-empty zone.
\end{theorem}

The right to left direction is straightforward, so we concentrate on
the opposite direction. 

\begin{lemma}\label{lem:a-post}
  For every zone $Z$, and transition $t$:
  \begin{equation*}
    Post(\Alu(Z),t) \subseteq \Alu(Post(Z,t))
  \end{equation*}
\end{lemma}

\begin{proof}
  Pick $\nu_1 \in Post(\Alu(Z),t)$. There exists a valuation $\nu \in
  \Alu(Z)$ such that $\nu \xrightarrow{t} \nu_1$. By definition of
  $\Alu$, there exists a valuation $\nu' \in Z$ such that $\nu' \lu
  \nu$. From Lemma \ref{lem:a-simulates}, $\nu' \xrightarrow{t}
  \nu_1'$ such that $\nu_1' \lu \nu_1$. Hence $\nu_1 \in
  \Alu(Post(Z,t))$. \qed
\end{proof}

The left to right implication of the theorem follows from the next
lemma and from the following invariant on the nodes of the tree that
is computed. For every node
$n=(q,Z,L,U)$:
\begin{enumerate}
\item if $n$ is $nottentative$, then $L, U$ are respectively the
  maximum of the $L_s, U_s$
  from all successor nodes $(q_s,Z_s,L_s, U_s)$ of $n$ (taking into account
  guards and clock resets, even if $Z_s$ is empty);
\item if $n$ is $tentative$ with respect to $(q',Z',L',U')$, then $L$ and $U$
  are equal to $L'$ and $U'$ respectively.
\end{enumerate}

\begin{lemma}\label{lem:LU-correct} 
  For every $(q,Z)$ reachable in $ZG(\Aa)$, there exists a 
  non tentative node $(q,Z_{1},L_1U_1)$ in the tree constructed by the
  LU-algorithm, such that $Z \subseteq \Alui(Z_1)$.
\end{lemma}
\begin{proof}
  The hypothesis is vacuously true for $(q_0, Z_0)$. Assume that the
  hypothesis is true for a node $(q,Z) \in ZG(\Aa)$. We prove that the
  lemma is true for every successor of $(q,Z)$.

  From hypothesis, there exists a node $(q,Z_1,L_1,U_1)$ in the tree
  constructed by the LU-algorithm such that $Z \subseteq
  \Alui(Z_{1})$. Let $t=(q, g, r, q')$ be a transition of $\Aa$ and
  let $(q,Z) \xrightarrow{t} (q',Z') \in ZG(\Aa)$. There are two
  cases.

  \paragraph{$(q,Z_1)$ is not tentative}

  Since $Z \subseteq \Alui(Z_{1})$, the transition $t$ is enabled from
  $\Alui(Z_{1})$. From Lemma~\ref{lem:a-simulates}, $t$ is enabled
  from $Z_{1}$ too. Since $Z \subseteq \Alui(Z_{1})$, we have
  $Post(Z,t) \subseteq Post(\Alui(Z_{1}),t)$, that is $Z' \subseteq
  Post(\Alui(Z_{1}),t)$. From Lemma~\ref{lem:a-post}, $Z' \subseteq
  \Alui(Post(Z_{1},t))$. We can take $Z_1'$ as $Post(Z_1,t)$ and then
  let $(q',Z_1', L_1'U_1')$ be the successor node in the tree computed by
  LU-algorithm. It remains to show that $Z' \subseteq \Alup(Z_1')$
  is the node corresponding to $(q',Z')$. This follows because by
  definition $L_1(x) \ge L_1'(x)$, $U_1(x) \ge U_1'(x)$ for all clocks
  $x$ that are not reset by the transition $t$ and for the clocks
  reset by $t$, $Z_1'$ entails $x=0$.

  \paragraph{$(q,Z_1)$ is tentative}

  If it is a tentative node, we know that there exists a non-tentative
  node $(q, Z_{2}, L_2U_2)$ in the tree constructed by the
  LU-algorithm such that $Z_{1} \subseteq \Closure_{L_2U_2}(Z_{2})$,
  that is, $Z_{1} \subseteq \mathfrak{a}_{L_2U_2}(Z_{2})$. The rest of
  the argument is the same as in the previous case with
  $(q,Z_2,L_2U_2)$ instead of $(q,Z_1, L_1U_1)$.

\qed
\end{proof}

%%% Local Variables: 
%%% mode: latex
%%% TeX-master: "m"
%%% End: 

\end{document}